\documentclass[aps,pre,reprint,amsmath,amssymb,floatfix,longbibliography]{revtex4-2}

\usepackage{graphicx}
\usepackage{bm}
\usepackage{natbib}
\usepackage{algpseudocode}
\usepackage{tikz}
\usetikzlibrary{arrows.meta,positioning,calc}

\tikzset{
  nd/.style   ={circle,fill=black,draw=black,inner sep=0pt,minimum size=3.4mm},
  ndg/.style  ={circle,fill=black!15,draw=black!15,inner sep=0pt,minimum size=3.4mm},
  foc/.style  ={circle,fill=white,draw=red,line width=0.9pt,inner sep=0pt,minimum size=3.4mm},
  ed/.style   ={draw=black,line width=0.5pt},
  edb/.style  ={draw=black,line width=1.0pt},
  edg/.style  ={draw=black!15,line width=0.5pt},
  msg/.style  ={draw=green!45!black,line width=0.9pt,-{Stealth[length=2.0mm]}},
  frame/.style={draw=black,line width=0.6pt},
  plab/.style ={font=\Large}
}
\def\ballcoords{
  \coordinate (i)  at (-0.55, 0.00);  \coordinate (j)  at ( 0.55, 0.06);
  \coordinate (u1) at (-1.55, 0.92);  \coordinate (u2) at (-1.62,-0.85);
  \coordinate (v1) at ( 1.58, 0.96);  \coordinate (v2) at ( 1.52,-0.88);
  \coordinate (w1) at (-2.52, 1.02);  \coordinate (w2) at (-1.68, 1.92);
  \coordinate (w3) at (-2.55,-0.92);  \coordinate (w4) at (-1.72,-1.86);
  \coordinate (w5) at ( 2.55, 1.08);  \coordinate (w6) at ( 1.70, 1.96);
  \coordinate (w7) at ( 2.52,-0.95);  \coordinate (w8) at ( 1.68,-1.90);
}

\newcommand{\E}{\mathbb{E}}
\newcommand{\Prob}{\mathbb{P}}
\newcommand{\Pstat}{\Prob_{\mathrm{stat}}}
\newcommand{\Pball}{\Prob_{\Theta}}
\newcommand{\one}{\mathbf{1}}

\newcommand{\xB}{\mathbf{x}}

\newcommand{\Fmap}{\mathcal{F}}

\begin{document}

\title{Space--time message passing for endemic diseases}

\author{Peter Mann}
\email{peter.mann@diai.io}
\affiliation{Data Insights AI, CodeBase Edinburgh, Argyle House, 3 Lady Lawson Street, Edinburgh, EH3 9DR, United Kingdom}

\author{Simon Dobson}
\affiliation{School of Computer Science, University of St Andrews, St Andrews, Fife KY16 9SX, United Kingdom }

\date{\today}

\begin{abstract}
Message passing is exact for progressive epidemics on trees: infection is unidirectional, and a node's neighbours are uncorrelated in the cavity graph. A recurrent disease breaks this even on a tree due to backtracking. A node infects a neighbour, recovers, and is reinfected by it, so infection traverses an edge in both directions. Unrolled along a time axis, the outward and return transmissions together with the node's own persistence in time form a closed cycle in the space--time graph, correlating the states that ordinary message passing assumes independent. We treat these correlations by solving the dynamics exactly inside a ball of radius $d$ about an edge and closing the ball's boundary with a single conditional per-edge message. For fixed-period susceptible--infected--susceptible dynamics this generates a hierarchy of closures on the endemic state indexed by $d$, in which the ball of radius $d$ treats exactly every space--time cycle of spatial reach at most $d$ and retains the remainder only through the mean rate supplied by its boundary. Linearising the resulting message map gives the endemic threshold, with an accuracy that increases with $d$. We compare the hierarchy against Monte Carlo simulation on random and empirical networks, finding excellent agreement.
\end{abstract}

\maketitle

\section{Introduction}\label{sec:intro}

Message passing is among the most powerful tools for quantitative calculation on
networks~\cite{10.1098/rspa.1935.0122,conf/aaai/Pearl82,Newman_2023,PhysRevE.107.054303,PhysRevE.82.016101,Cantwell_Newman_2019}. It enjoys success across a wide variety of applications, including statistical physics \cite{doi:10.1126/sciadv.abf1211}, epidemiology \cite{PhysRevE.82.016101,PhysRevE.111.064301}, Bayesian inference, graphical models \cite{jordan1998learning}, signal processing \cite{gallager1963low, NIPS1997_0245952e} and NP-hard computations on graphs \cite{mezard2002analytic}. The quantities of interest are abstractly modelled as probabilistic \emph{states} on each node. Self-consistent equations are written among the neighbours and solved by iteration to a stable fixed point. The power of message passing rests on a single property of trees: removing a node splits its neighbours into disjoint subtrees; hence the states of neighbours are \emph{independent} and the equations converge. The method is exact on trees, and works well on ``locally tree-like'' networks that contain only long-range loops. It fails when two neighbours are joined by a short loop, because the loop correlates the states the message passing equations assumes are independent \cite{Yedidia2001Generalized, PhysRevE.107.054303}. Cantwell and Newman~\cite{Cantwell_Newman_2019} removed this restriction by working with neighbourhoods: one enlarges the region treated exactly until it
contains the short loops, passes messages between neighbourhoods on their
boundaries, and obtains a controlled series of approximations whose $r$th member is exact when the network has no cycle longer than $r+2$. Standard
message passing is the $r=0$ member~\cite{PhysRevE.82.016101}.

What exact message passing on trees has so far required is that the dynamics be
\emph{progressive}: that each node's state change in one direction only. Consider the \emph{susceptible--infected--removed} (SIR) process, in which a disease
spreads through the network. Infection flows outward from a node and never
returns, so the cavity construction removes every feedback path and the infection states of the neighbours are uncorrelated. Message passing for such processes may be static, computing final outbreak sizes, or fully
dynamic~\cite{PhysRevE.82.016101,Lokhov2015Dynamic,Zhang_2012}, tracking the probability that each node is infectious at each time; in either case the progressive, unidirectional structure is what makes it exact on a tree.

In a recurrent \emph{susceptible--infected--susceptible} (SIS) process infection
is allowed to return to a given
node~\cite{Harris_1974,PhysRevLett.104.258701,PhysRevLett.116.258301,Shrestha_Scarpino_Moore_2015,PhysRevLett.107.068701,Castellano_PastorSatorras_2018,wg82-f4lf,385j-2f29,PhysRevX.3.021004}, so the same edge carries infection in both directions. We will show that the outward transmission and its later return close a cycle in \emph{space--time}, and that recurrent dynamics therefore demand an extension of the Cantwell--Newman neighbourhood machinery even on networks containing no \emph{spatial} loops at all. These cycles are the analogue, for a recurrent disease, of the obstruction that spatial loops generate for ordinary message passing.

The remedy we propose is a neighbourhood construction in space \emph{and} time. We treat a ball of fixed radius about an edge exactly, so that it contains the returning walks of bounded spatial reach, and close its boundary with a single per-edge message. We carry this out for fixed-period SIS, in which a deterministic infectious period $\tau$ keeps every ball finite and hence exactly solvable \cite{PhysRevLett.104.258701}. The result is a hierarchy of closures for the endemic state ordered by the radius of the ball, and linearising the construction delivers the endemic threshold at every depth.

The paper is organised as follows. Section~\ref{sec:model} outlines the model and Sec.~\ref{sec:cycles} identifies the obstruction. Section~\ref{sec:ball}
constructs the ball, the message map it defines, and the endemic threshold that follows from linearising it. Section~\ref{sec:results} compares with simulation and Sec.~\ref{sec:discussion} relates the construction to other treatments of the same correlation.

\section{Model}\label{sec:model}
The substrate is a configuration-model graph with degree distribution
$\{p_k\}$, locally tree-like, with excess-degree generating function
\begin{equation}\label{eq:gf}
  G_1(z)=\frac{\sum_k k\,p_k\,z^{k-1}}{\langle k\rangle},
\end{equation}
so that $G_1'(1)=\langle k(k-1)\rangle/\langle k\rangle$; the random $k$-regular
graph (RRG) has $G_1(z)=z^{k-1}$ and $G_1'(1)=k-1$~\cite{Newman_2019,
Dorogovtsev_F._2022}.

Each node carries an age $x\in\{0,1,\dots,\tau\}$. The value $x=0$ denotes a susceptible node (S); a value $x=m$ with $1\le m\le\tau$ denotes an infectious
node (I) with $m$ steps remaining before recovery. A synchronous update applies,
in order,
\begin{enumerate}
  \item[(i)] \emph{transmission}: each infectious node infects each susceptible
    neighbour with probability $r$, so a susceptible node with $n$ infectious
    neighbours is infected with probability $1-(1-r)^n$;
  \item[(ii)] \emph{ageing}: every infectious age decreases by one, with age $1$
    recovering to $0$;
  \item[(iii)] \emph{onset}: newly infected nodes are set to age $\tau$.
\end{enumerate}
Throughout, ``update'' means one application of (i)--(iii), and time is measured
in updates. The order within an update matters. Transmission in step~(i) is
evaluated on the configuration as it stands at the start of the update, so all
transmissions within an update are simultaneous and a node infected during
update $t$ cannot itself transmit until update $t+1$; it is then infectious for
exactly the $\tau$ updates $t+1,\dots,t+\tau$, and is susceptible again at
$t+\tau+1$. This simultaneity is what allows the joint dynamics of a region to
factorise into independent per-node moves once the current configuration is
fixed, a property we use in Sec.~\ref{sec:transfer}.

The deterministic infectious period is what makes the construction below finite.
A node's entire memory of the epidemic is the single integer $x$, so the state
space of any bounded region is finite and its stationary distribution can be
computed exactly.

All probabilities below are evaluated in the endemic steady state, written $\Pstat$. On a finite network the process is absorbing and its invariant law is the empty configuration, so $\Pstat$ denotes the quasi-stationary law, conditioned on survival, which is what Algorithm~3 estimates and which becomes a genuine stationary law as $N\to\infty$ above the threshold. The observable is the stationary prevalence $\rho=\Pstat(\text{a node is infectious})$.

\section{Space--time cycles}\label{sec:cycles}

\begin{figure}[!t]
\centering
\begin{tikzpicture}[scale=0.9, every node/.style={scale=0.8}]

\begin{scope}[shift={(0,0)}]
  \draw[frame] (-2.7,-2.3) rectangle (2.7,2.3);
  \coordinate (i) at (-0.80,0);  \coordinate (j) at (0.80,0);
  \draw[ed] (i) -- ++(145:1.35);  \draw[ed] (i) -- ++(215:1.35);
  \draw[ed] (j) -- ++(35:1.35);   \draw[ed] (j) -- ++(325:1.35);
  \coordinate (a1) at ($(i)+(145:1.35)$);  \coordinate (a2) at ($(i)+(215:1.35)$);
  \coordinate (b1) at ($(j)+(35:1.35)$);   \coordinate (b2) at ($(j)+(325:1.35)$);
  \draw[ed] (a1) -- ++(110:0.85); \draw[ed] (a1) -- ++(180:0.85);
  \draw[ed] (a2) -- ++(250:0.85); \draw[ed] (a2) -- ++(180:0.85);
  \draw[ed] (b1) -- ++(70:0.85);  \draw[ed] (b1) -- ++(0:0.85);
  \draw[ed] (b2) -- ++(290:0.85); \draw[ed] (b2) -- ++(0:0.85);
  \draw[ed] (i) -- (j);
  \foreach \p in {a1,a2,b1,b2} \node[nd] at (\p) {};
  \draw[line width=1.1pt,draw=orange!85!black,-{Stealth[length=2.4mm]}]
        ($(i)+(0.16,0.14)$) to[bend left=45] ($(j)+(-0.16,0.14)$);
  \draw[line width=1.1pt,draw=orange!85!black,-{Stealth[length=2.4mm]}]
        ($(j)+(-0.16,-0.14)$) to[bend left=45] ($(i)+(0.16,-0.14)$);
  \node[font=\footnotesize,orange!70!black] at (0,0.78) {infect};
  \node[font=\footnotesize,orange!70!black] at (0,-0.80) {reinfect};
  \node[foc] at (i) {};  \node[foc] at (j) {};
  \node[font=\footnotesize] at ($(i)+(-0.30,0.28)$) {$i$};
  \node[font=\footnotesize] at ($(j)+( 0.30,0.28)$) {$j$};
  \node[plab] at (2.32,-1.94) {(a)};
\end{scope}

\begin{scope}[shift={(0,-5.0)}]
  \draw[frame] (-2.7,-2.3) rectangle (2.7,2.3);
  \foreach \r/\y in {0/-1.35, 1/0.10, 2/1.55} {
    \coordinate (I\r) at (-0.30,\y);  \coordinate (J\r) at (1.55,\y);
  }
  \draw[-{Stealth[length=2.2mm]},draw=black!50,line width=0.5pt] (-2.42,-1.62) -- (-2.42,1.85);
  \node[rotate=90,font=\footnotesize,black!50] at (-2.62,0.15) {time};
  \foreach \r/\lab in {0/$t$,1/$t{+}1$,2/$t{+}2$}
    \node[font=\footnotesize,black!50,anchor=east] at ($(I\r)+(-0.42,0)$) {\lab};
  \draw[edg,-{Stealth[length=1.7mm]}] (J0) -- (I1);
  \draw[edg,-{Stealth[length=1.7mm]}] (I1) -- (J2);
  \draw[edg,-{Stealth[length=1.7mm]}] (J0) -- (J1);
  \draw[edg,-{Stealth[length=1.7mm]}] (J1) -- (J2);
  \draw[draw=blue!70!black,line width=1.5pt,-{Stealth[length=2.4mm]}] (I0) -- (I1);
  \draw[draw=blue!70!black,line width=1.5pt,-{Stealth[length=2.4mm]}] (I1) -- (I2);
  \draw[draw=orange!85!black,line width=1.5pt,-{Stealth[length=2.4mm]}] (I0) -- (J1);
  \draw[draw=orange!85!black,line width=1.5pt,-{Stealth[length=2.4mm]}] (J1) -- (I2);
  \foreach \r in {0,1,2} { \node[nd] at (J\r) {}; }
  \node[nd] at (I1) {};
  \node[foc] at (I0) {};  \node[foc] at (I2) {};
  \node[font=\footnotesize] at (-0.30,-1.95) {$i$};
  \node[font=\footnotesize] at ( 1.55,-1.95) {$j$};
  \node[font=\footnotesize,anchor=west,orange!80!black] at (0.72,1.02) {transmission};
  \node[font=\footnotesize,anchor=east,blue!70!black]   at (-0.48,0.82) {persistence};
  \node[plab] at (2.32,-1.94) {(b)};
\end{scope}
\end{tikzpicture}
\caption{Reinfection is a cycle in space--time. \emph{(a)}~On the spatial graph infection passes from $i$ to $j$ and later returns, carrying transmission in both directions. \emph{(b)}~The same two events unrolled along a time axis. The space--time graph $\mathcal{G}$ carries a \emph{persistence} edge $(v,t)\to(v,t+1)$ at every node $v$, which advances that node's age, and a \emph{transmission} edge $(u,t)\to(v,t+1)$ along every spatial edge $(u,v)$; faint arrows show the edges not involved here. The
transmission path $(i,t)\to(j,t+1)\to(i,t+2)$ (orange) and the persistence path
$(i,t)\to(i,t+1)\to(i,t+2)$ (blue) are two distinct paths between the
same pair of space--time nodes (open, red), and therefore close an undirected cycle of length four.}
\label{fig:echo}
\end{figure}
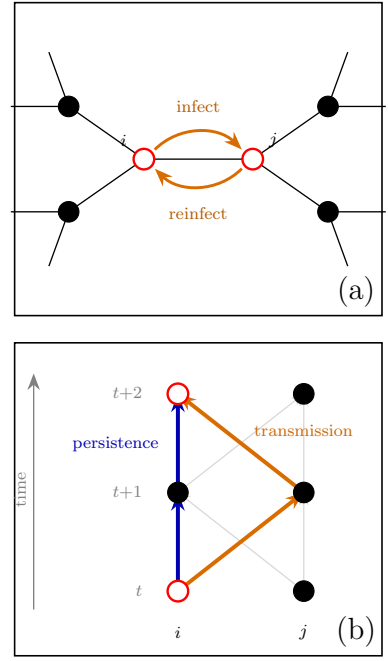

The assumption behind message passing is that the arrivals of infection on
distinct edges of a node are independent, and that the far endpoints of those
edges are not coupled through any other path. For a progressive epidemic on a
tree this holds automatically: infection leaves a node once and never comes back,
so distinct neighbours act through disjoint histories. 

Recurrence breaks this assumption in time even when the graph remains a tree in space. To see this, we unroll the dynamics along a time axis. The \emph{space--time graph}
$\mathcal{G}$ has a node $(v,t)$ for every node $v$ and every update $t$, and
two kinds of directed edge (Fig.~\ref{fig:echo}): a \emph{persistence} edge
$(v,t)\to(v,t+1)$ at every node, which carries that node's age forward, and a
\emph{transmission} edge $(u,t)\to(v,t+1)$ for every spatial edge $(u,v)$. Every
edge of $\mathcal{G}$ points forward in time, so $\mathcal{G}$ is acyclic as a
directed graph; it nonetheless contains undirected cycles, and it is these that
obstruct message passing.

Message passing requires that there is a \emph{unique path} between the nodes whose states it treats as independent; two distinct paths between the same pair of nodes
correlate their endpoints. In $\mathcal{G}$ this happens as soon as
infection returns. The out-and-back transmission
\begin{equation}\label{eq:echo}
  (i,t)\to(j,t+1)\to(i,t+2)
\end{equation}
and the persistence path $(i,t)\to(i,t+1)\to(i,t+2)$ are two distinct directed
paths between $(i,t)$ and $(i,t+2)$. Together they close an undirected cycle of
length four in $\mathcal{G}$, and the states at its endpoints are correlated
exactly as a spatial loop would correlate them in percolation. We call this shortest
cycle the \emph{echo}, after Ref.~\cite{Shrestha_Scarpino_Moore_2015}. In this way, recurrence produces a correlation between the ages of $i$ and $j$: the joint
distribution of $(x_i,x_j)$ fails to factorise. The cycle is generated
by the timing of infection and recovery, while its size is measured by the
spatial reach defined next.

Because the construction below is centred on an edge rather than on a node, we
measure distance from the central edge throughout, writing
\begin{equation}\label{eq:dist}
  \mathrm{dist}\big(v,(i,j)\big)=\min\{\mathrm{dist}(v,i),\,\mathrm{dist}(v,j)\},
\end{equation}
and define the \emph{spatial reach} of a space--time cycle as the largest such
distance attained by any node of its spatial trace. The echo has reach $0$, its
trace never leaving the two endpoints; an excursion that travels out to distance
$L$ and returns closes a cycle of reach $L$. Cycle lengths are unambiguous here
because every edge of $\mathcal{G}$ advances time by exactly one update, so any
two paths between the same pair of space--time nodes contain the same number of
edges.

Equivalently, in the language of the percolation literature, the cavity
construction is the statement that the walks propagating infection are
\emph{non-backtracking}. For a recurrent disease that restriction is false, and
the backtracking walk $i\to j\to i$ is the leading contribution a non-backtracking closure
omits~\cite{Castellano_PastorSatorras_2018}. The echo of \eqref{eq:echo} is that
walk drawn in space--time, and the construction of Sec.~\ref{sec:ball} is a
controlled reintroduction of backtracking, ordered by spatial reach.

In ordinary message passing it is typical to enumerate the set of reachable nodes to determine the size of the giant connected component in the network \cite{PhysRevE.76.045101,PhysRevE.82.016101}. 
The space--time graph also
suggests a percolation-like reachable set, but for the fixed-period SIS process
that set is not the same as being infectious. To see this, realise the
transmission step of Sec.~\ref{sec:model} as a fixed transmission field Ref.~\cite{Harris_1974}. Attach to
every ordered edge $u\to v$ and every update $t$ an independent Bernoulli
variable $\eta_{u\to v}(t)$ of success probability $r$, and let $u$ transmit to
$v$ on update $t$ when $u$ is infectious and $\eta_{u\to v}(t)=1$. Averaging over
these variables returns rule~(i) exactly, and holding them fixed lets two dynamics be
compared within each realisation. Define the arrival indicator
\begin{equation}\label{eq:chi1}
  \chi_v(t\mid z)=1-\prod_{u\sim v}\big(1-\eta_{u\to v}(t)\,\one[z_u(t)\ge1]\big),
\end{equation}
equal to one exactly when at least one neighbour that is infectious in the
process $z$ transmits to $v$ on update $t$; rule~(i) of Sec.~\ref{sec:model}
infects a susceptible $v$ precisely when $\chi_v(t\mid x)=1$. Couple SIS in this
way to an ungated reachability process $y_v(t)\in\{0,\ldots,\tau\}$, obtained by
dropping the requirement that $v$ be susceptible,
\begin{equation}\label{eq:ungated}
y_v(t+1)=
\begin{cases}
\tau, & \chi_v(t\mid y)=1,\\
\max\{y_v(t)-1,0\}, & \text{otherwise}.
\end{cases}
\end{equation}

Unlike SIS, an arriving transmission resets the clock whenever it arrives, so the
update \eqref{eq:ungated} is monotone in both the current ages and the transmission
field. The SIS update is monotone in neither, because a transmission reaching an
already infectious node is void rather than prolonging: with $\tau=2$ and $v$
susceptible at $t=0$, a single transmission into $v$ at $t=1$ gives $x_v(2)=2$ and
$x_v(3)=1$, whereas adding a second transmission at $t=0$ gives $x_v(1)=2$, voids the
transmission at $t=1$, and leaves $x_v(3)=0$; a transmission has been added and the
event $\{x_v(3)\ge1\}$ removed. It is this asymmetry that makes the comparison
run in one direction only. If
$y(0)=x(0)$, then the coupling preserves the stronger age-wise order
$x_v(t)\le y_v(t)$ for all $v$ and $t$: the countdown case needs the age
inequality itself, while a reset in $y$ only increases its age. Hence
\begin{equation}\label{eq:reachbound}
  \{x_i(t)\ge1\}\subseteq\{y_i(t)\ge1\}.
\end{equation}
The process $y$ is the genuine oriented-percolation object: $y_i(t)\ge1$ if
and only if there is an $\eta$-open oriented path in $\mathcal{G}$ from the
initial infected set to $(i,t)$, with gaps between successive transmissions no
longer than the persistence window $\tau$. Deterministic-period SIS is bounded
by this cluster event but is not itself one: by the example above,
$\one[x_i(T)\ge1]$ is not an increasing function of the transmission field, and this
non-attractiveness is the same obstruction that prevents a static percolation
convolution from computing prevalence. The comparison diagnoses that obstruction
rather than providing a tight approximation: under persistent exposure the reachable process
remains occupied, whereas a fixed-period SIS node still spends one update in
every $\tau+1$ susceptible on its deterministic cycle.

A related but distinct correlation acts between two neighbours $j$ and $j'$ of a
common node $i$: while $i$ is infectious it drives both at once, so their states
are correlated through their shared ancestor $(i,t)$. This is a fork in
$\mathcal{G}$ rather than a cycle, and we call it a \emph{common-driver}
correlation. It is resolved at depth $1$ of the hierarchy, where the full
neighbourhood of each endpoint is retained.

\section{The ball}\label{sec:ball}

\begin{figure*}[!t]
\centering
\resizebox{\textwidth}{!}{%
\begin{tikzpicture}[scale=0.90]
\foreach \px/\py/\lab in {0/0/(a), 8.2/0/(b), 16.4/0/(c), 24.6/0/(d)} {
  \begin{scope}[shift={(\px,\py)}]
    \draw[frame] (-3.3,-2.7) rectangle (3.3,2.7);
    \node[plab] at (2.90,-2.32) {\lab};
  \end{scope}
}

\begin{scope}[shift={(0,0)}] \ballcoords
  \draw[ed] (i)--(u1) (i)--(u2) (j)--(v1) (j)--(v2);
  \draw[ed] (u1)--(w1) (u1)--(w2) (u2)--(w3) (u2)--(w4);
  \draw[ed] (v1)--(w5) (v1)--(w6) (v2)--(w7) (v2)--(w8);
  \foreach \p/\A/\B in {w1/118/198, w2/91/171, w3/160/240, w4/187/267,
                        w5/-17/63, w6/9/89, w7/-61/19, w8/-88/-8} {
    \draw[ed] (\p) -- ++(\A:0.62);  \draw[ed] (\p) -- ++(\B:0.62); }
  \draw[ed] (i)--(j);
  \foreach \p in {u1,u2,v1,v2,w1,w2,w3,w4,w5,w6,w7,w8} \node[nd] at (\p) {};
  \node[foc] at (i) {}; \node[foc] at (j) {};
\end{scope}
\begin{scope}[shift={(8.2,0)}] \ballcoords
  \draw[edg] (i)--(u1) (i)--(u2) (j)--(v1) (j)--(v2);
  \draw[edg] (u1)--(w1) (u1)--(w2) (u2)--(w3) (u2)--(w4);
  \draw[edg] (v1)--(w5) (v1)--(w6) (v2)--(w7) (v2)--(w8);
  \foreach \p/\A/\B in {w1/118/198, w2/91/171, w3/160/240, w4/187/267,
                        w5/-17/63, w6/9/89, w7/-61/19, w8/-88/-8} {
    \draw[edg] (\p) -- ++(\A:0.62);  \draw[edg] (\p) -- ++(\B:0.62); }
  \foreach \p in {u1,u2,v1,v2,w1,w2,w3,w4,w5,w6,w7,w8} \node[ndg] at (\p) {};
  \draw[edb] (i)--(j);
  \foreach \p in {u1,u2} \draw[msg] ($(\p)!0.55!(i)$) -- ($(\p)!0.86!(i)$);
  \foreach \p in {v1,v2} \draw[msg] ($(\p)!0.55!(j)$) -- ($(\p)!0.86!(j)$);
  \node[foc] at (i) {}; \node[foc] at (j) {};
  \node[font=\large,green!45!black] at (0.05,-1.30) {$\sigma$};
\end{scope}
\begin{scope}[shift={(16.4,0)}] \ballcoords
  \draw[edg] (u1)--(w1) (u1)--(w2) (u2)--(w3) (u2)--(w4);
  \draw[edg] (v1)--(w5) (v1)--(w6) (v2)--(w7) (v2)--(w8);
  \foreach \p/\A/\B in {w1/118/198, w2/91/171, w3/160/240, w4/187/267,
                        w5/-17/63, w6/9/89, w7/-61/19, w8/-88/-8} {
    \draw[edg] (\p) -- ++(\A:0.62);  \draw[edg] (\p) -- ++(\B:0.62); }
  \foreach \p in {w1,w2,w3,w4,w5,w6,w7,w8} \node[ndg] at (\p) {};
  \draw[edb] (i)--(j) (i)--(u1) (i)--(u2) (j)--(v1) (j)--(v2);
  \foreach \a/\b in {w1/u1,w2/u1,w3/u2,w4/u2,w5/v1,w6/v1,w7/v2,w8/v2}
    \draw[msg] ($(\a)!0.42!(\b)$) -- ($(\a)!0.80!(\b)$);
  \foreach \p in {u1,u2,v1,v2} \node[nd] at (\p) {};
  \node[foc] at (i) {}; \node[foc] at (j) {};
\end{scope}
\begin{scope}[shift={(24.6,0)}] \ballcoords
  \draw[edb] (i)--(j) (i)--(u1) (i)--(u2) (j)--(v1) (j)--(v2);
  \draw[edb] (u1)--(w1) (u1)--(w2) (u2)--(w3) (u2)--(w4);
  \draw[edb] (v1)--(w5) (v1)--(w6) (v2)--(w7) (v2)--(w8);
  \foreach \p/\A/\B in {w1/118/198, w2/91/171, w3/160/240, w4/187/267,
                        w5/-17/63, w6/9/89, w7/-61/19, w8/-88/-8} {
    \draw[msg] ($(\p)+(\A:0.66)$) -- ($(\p)+(\A:0.26)$);
    \draw[msg] ($(\p)+(\B:0.66)$) -- ($(\p)+(\B:0.26)$); }
  \foreach \p in {u1,u2,v1,v2,w1,w2,w3,w4,w5,w6,w7,w8} \node[nd] at (\p) {};
  \node[foc] at (i) {}; \node[foc] at (j) {};
\end{scope}
\end{tikzpicture}}
\caption{Constructing the ball. \emph{(a)}~A patch of a locally tree-like network of degree $k=3$. The construction is centred on the \emph{edge} $(i,j)$, whose endpoints are drawn open and red; this is the one structural difference from Ref.~\cite{Cantwell_Newman_2019}, where the neighbourhood is centred on a node. \emph{(b)}~The depth-$0$ ball $B_0=\{i,j\}$ treats only the two endpoints exactly and closes each of their four remaining edges with the boundary message $\sigma$ (green). It is the smallest ball containing the echo of Fig.~\ref{fig:echo}. \emph{(c)}~The depth-$1$ ball $B_1$ adds every neighbour of both endpoints, moving the boundary out by one shell. \emph{(d)}~The depth-$2$ ball $B_2$ adds a further shell. At each depth the nodes drawn black are treated exactly and everything grey is represented only through those boundary messages; enlarging the ball resolves space--time cycles of greater spatial reach at the cost of a state space $(\tau+1)^{|B_d|}$ that grows doubly exponentially with $d$ on a regular tree.}
\label{fig:ball}
\end{figure*}
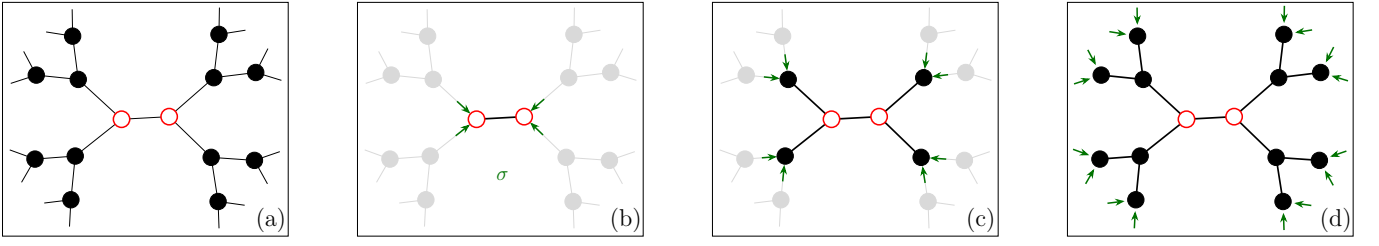

The construction is to enlarge the region treated exactly until it holds the correlations of interest, solve its dynamics in full, and represent the rest of the network by a single message on its boundary. Letting the region grow generates a hierarchy ordered by its radius. This is similar in spirit to the Cantwell-Newman neighbourhood approach for resolving spatial loops \cite{Cantwell_Newman_2019}.

We build it from the \emph{ball} $B_d(i,j)$ of an edge: the edge $(i,j)$ together
with every node within distance $d$ of it, and every edge among them
(Fig.~\ref{fig:ball}). We use ``ball'' for this object throughout and reserve
``neighbourhood'' for the construction of Ref.~\cite{Cantwell_Newman_2019} from
which it is adapted.

We first define the \emph{message}. Message passing represents everything outside the region treated exactly by a single number crossing its boundary, and that number lives on a \emph{directed} edge. We write
\begin{equation}\label{eq:msg}
  \sigma_{i\leftarrow j}
  =\Pstat\big(j \text{ infectious }
    \mid i \text{ susceptible}\big)
\end{equation}
for the infectiousness of $j$ as seen along the edge $j\to i$ at a given update. The quantity $\sigma_{i\leftarrow j}$ describes $j$, with its own degree and its own neighbourhood behind it, while $\sigma_{j\leftarrow i}$ describes $i$, and on a general graph the two differ. A network with $E$ edges therefore carries $2E$ messages, one for each direction of each edge. The conditioning in \eqref{eq:msg} is important: the message supplies the rate at which a
\emph{susceptible} node is reinfected from outside the region, and for a
recurrent disease replacing $\sigma$ by its unconditional counterpart discards
the echo on that edge. Section~\ref{sec:resum} shows why the conditional
probability is the object the closure must match.

The \emph{message map} supplies the equations the messages satisfy. The construction has two nested iterations. The inner iteration is the epidemic. Fix a directed edge, build the ball $B_d(i,j)$ about it, supply a message on every edge leaving the ball, and run the resulting finite Markov chain to stationarity; reading the conditional infectiousness off the central edge then returns an updated message,
\begin{equation}\label{eq:Fmap}
  \sigma'_{i\leftarrow j}
  =\Fmap_d\big[\{\sigma_{v\leftarrow w}\}\big],
\end{equation}
where $(v,w)$ runs over the edges leaving $B_d(i,j)$. All of this is a single evaluation of $\Fmap_d$, and the rest of this section specifies its steps.

The outer iteration is the search for consistency. There is one equation \eqref{eq:Fmap} for each of the $2E$ directed edges, and iterating the system until the messages stop moving gives the model at depth $d$ as its simultaneous fixed point.

\subsection{A single driven node}\label{sec:atom}

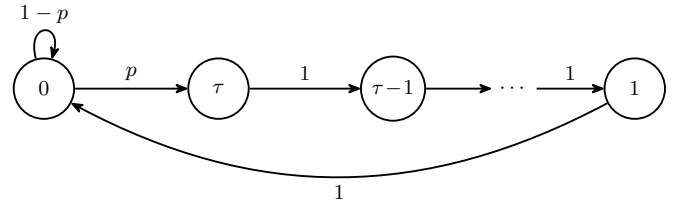
\begin{figure}[t]
\centering
\resizebox{\columnwidth}{!}{%
\begin{tikzpicture}[>={Stealth[round]},thick,every node/.style={font=\small}]
  \node[circle,draw,minimum size=9mm] (s)  at (0,0)   {$0$};
  \node[circle,draw,minimum size=9mm] (t)  at (2.6,0) {$\tau$};
  \node[circle,draw,minimum size=9mm] (tm) at (5.2,0) {$\tau\!-\!1$};
  \node                               (d)  at (7.0,0) {$\cdots$};
  \node[circle,draw,minimum size=9mm] (o)  at (8.8,0) {$1$};
  \draw[->] (s)  -- node[above]{$p$} (t);
  \draw[->] (t)  -- node[above]{$1$} (tm);
  \draw[->] (tm) -- (d);
  \draw[->] (d)  -- node[above]{$1$} (o);
  \draw[->] (o)  to[bend left=28] node[below]{$1$} (s);
  \draw[->] (s)  to[loop above,looseness=6] node[above]{$1-p$} (s);
\end{tikzpicture}}
\caption{The age chain \eqref{eq:Kkernel} of a single node driven by reinfection
probability $p$. Infection carries the node from the susceptible state $0$ to age
$\tau$; ageing then walks it deterministically down to $0$ again. Every
transition other than the self-loop is one-way, so the chain is irreversible:
probability circulates around the cycle and the reverse transitions have
probability zero.}
\label{fig:chain}
\end{figure}

Every node in the ball evolves by the same microscopic rules. Consider a node with reinfection probability $p$. From the update rules of Sec.~\ref{sec:model} its age is a Markov chain on $\{0,\dots,\tau\}$ with kernel
\begin{equation}\label{eq:Kkernel}
  K(x\!\to\! x'\mid p)=
  \begin{cases}
   p\,\delta_{x',\tau}+(1-p)\,\delta_{x',0}, & x=0,\\[2pt]
   \delta_{x',x-1}, & x\ge1 .
  \end{cases}
\end{equation}
A susceptible node is reinfected with probability $p$, in which case it jumps to
age $\tau$, and otherwise stays susceptible; an infectious node ages
deterministically, reaching $x'=0$ from $x=1$. The chain is therefore a single
directed cycle carrying one self-loop, Fig.~\ref{fig:chain}.

\subsubsection{The stationary distribution}\label{sec:pi}

The kernel describes one update \cite{Norris_1997,levin2008markov}. What the construction needs is the node's
behaviour over many updates, and that is carried by the chain's stationary
distribution
\begin{equation}\label{eq:pidef}
  \pi=(\pi_0,\pi_1,\dots,\pi_\tau),\qquad
  \pi_x\ge0,\quad \sum_{x=0}^{\tau}\pi_x=1 .
\end{equation}
The node has $\tau+1$ distinguishable states and we need to know how much of its time it spends in each one. The component $\pi_x$ is the long-run fraction of updates on which the node
has age $x$: thus $\pi_0$ is the fraction of time the node is susceptible, and
$\pi_m$ for $m\ge1$ the fraction of time it is infectious with exactly $m$
updates remaining before recovery. The probability that the node is infectious is $\sum_{m\ge1}\pi_m$, which discards the age information and retains only the susceptible--infectious distinction.

\subsubsection{Balance}\label{sec:balance}

The distribution is stationary when one further update leaves it unchanged,
$\pi K=\pi$. Written out for a single age $x'$ this reads
$\pi_{x'}=\sum_{x}\pi_x K(x\to x')$: the weight arriving on $x'$ from every state
that can reach it must equal the weight $x'$ already carried. These are the
\emph{balance equations}, and for the kernel \eqref{eq:Kkernel} they are
\begin{align}\label{eq:balance}
  \pi_\tau =&\ p\,\pi_0,\\
  \pi_m =&\ \pi_{m+1}\quad (1\le m\le\tau-1),\label{eq:balance2}\\
  \pi_0 =&\ (1-p)\,\pi_0+\pi_1 .\label{eq:balance3}
\end{align}
Each line records the only ways of entering the state on its left: age $\tau$ can
be reached only by a fresh infection, an intermediate age $m$ only by ageing from
$m+1$, and the susceptible state either by failing to be infected or by recovering
from age $1$.

Two observations then solve \eqref{eq:balance}--\eqref{eq:balance3} by
inspection. First, consider the
boundary separating the susceptible state from the infectious block
$\{1,\dots,\tau\}$. In the stationary state the probability crossing that
boundary per update must be equal in the two directions, since otherwise weight
would accumulate on one side of it. A node leaves the susceptible state with
probability $p$, contributing flux $p\,\pi_0$; it returns only from age $1$, and
does so with certainty, contributing flux $\pi_1$. The last line of
\eqref{eq:balance}, rearranged, is exactly this statement, $p\,\pi_0=\pi_1$. Second, the
infectious ages form a deterministic pipeline: a node entering at age $\tau$
passes through each of $\tau,\tau-1,\dots,1$ exactly once, so each of these
states is entered only from its predecessor and left only to its successor, both
with certainty. Balance at each therefore equates neighbouring weights, and all
$\tau$ infectious ages carry the same occupancy,
\begin{equation}\label{eq:piflat}
  \pi_1=\pi_2=\dots=\pi_\tau=p\,\pi_0 .
\end{equation}
Normalisation fixes the remaining scale through $\pi_0(1+\tau p)=1$, and the
infectious occupancy follows,
\begin{equation}\label{eq:renewal}
  \sum_{m=1}^{\tau}\pi_m=\frac{\tau p}{1+\tau p}
  =\frac{\tau}{\tau+\langle W\rangle},
\end{equation}
where $\langle W\rangle=1/p$ is the mean number of consecutive updates spent
susceptible. The second form is the renewal reading: the node is infectious for a
fixed window $\tau$ out of every cycle of mean length $\tau+\langle W\rangle$, so
the infectious occupancy is the ratio of the two.

For a fixed reinfection probability $p$, $\pi$ is the single-node stationary law
from which the infectious occupancy \eqref{eq:renewal} is read. For $0<p<1$ the chain is irreducible, every age being reachable from every other, and aperiodic, the self-loop at age $0$ destroying the periodicity the bare cycle would impose. We remark that at $p=1$ the chain is periodic and reducible at $p=0$. Exactly one $\pi$ satisfies $\pi K=\pi$, and any initial distribution converges to it, meaning it is a unique and attracting fixed-point.

\subsection{The transfer operator}\label{sec:transfer}

The single-node kernel is assembled into a chain on the whole ball in two steps:
first the edges leaving the ball are converted into a reinfection probability,
then the per-node kernels are multiplied together.

\subsubsection{Closing the boundary}\label{sec:hazard}

Let us name the edges leaving the ball its \emph{background} edges. A
background edge $(v,w)$ delivers infection on a given update only if $w$ is
infectious and transmission across it succeeds; given that $v$ is susceptible,
the closed boundary assigns that delivery probability
$r\,\sigma_{v\leftarrow w}$. Distinct background edges of $v$ open onto
disjoint subtrees. The independence used in the product below is the
full-history conditional independence made explicit in Sec.~\ref{sec:resum}, so
$v$ escapes reinfection from outside the ball only if every one of them fails:
\begin{equation}\label{eq:bgsurvival}
  \Prob\big(\text{no background edge of }v\text{ transmits}\big)
  =\!\!\prod_{\substack{w\notin B\\ w\sim v}}\!\!\big(1-r\,\sigma_{v\leftarrow w}\big),
\end{equation}
where $w\notin B$ indicates nodes not in the ball and $w\sim v$ indicates that $w$ is a neighbour of $v$.

\subsubsection{Assembling the chain}\label{sec:assemble}

Within the ball the process is a finite Markov chain. Its state is the joint
age configuration $\xB=(x_v)_{v\in B}$; one update carries $\xB$ to a successor
$\xB'$, and the object we want is the chain's stationary distribution $\Theta(\xB)$,
the long-run probability of each configuration. Collect the one-update transition
probabilities into the \emph{transfer operator} $T_B$, whose entry
$T_B[\xB\to\xB']$ is the probability that the ball moves from $\xB$ to $\xB'$ in
a single update, and of which $\Theta$ is the invariant distribution.

Because transmission travels along each edge independently, once the current
configuration is fixed the nodes update independently of one another: a node's
next age depends only on its own age and on which of its neighbours are currently
infectious. The transfer operator therefore factorises into a product of the
single-node kernels \eqref{eq:Kkernel},
\begin{equation}\label{eq:TB}
  T_B[\xB\to\xB']=\prod_{v\in B}K\big(x_v\to x_v'\mid p_v(\xB)\big),
\end{equation}
where $p_v$ is node $v$'s own reinfection probability. This conditional
independence is what makes the ball tractable: the joint chain is large, but it
is built from $|B|$ copies of a $(\tau{+}1)$-state kernel.

The reinfection probability $p_v(\xB)$ is the chance that node $v$, if
susceptible, is infected during the update given the configuration $\xB$. It gathers every edge that can carry infection to $v$, and these fall into two kinds. The neighbours resolved inside the ball have their infectious states read directly from $\xB$, each contributing a survival factor $1-r$ when infectious and $1$ otherwise. The neighbours outside the ball are not resolved and contribute the messages of \eqref{eq:bgsurvival}. A susceptible $v$ stays susceptible only if none of these edges transmits, so
\begin{equation}\label{eq:pv}
  p_v(\xB)=1-\!\!\prod_{\substack{u\in B\\ u\sim v}}\!\!\big(1-r\,\one[x_u\ge1]\big)
           \!\!\prod_{\substack{w\notin B\\ w\sim v}}\!\!\big(1-r\,\sigma_{v\leftarrow w}\big).
\end{equation}
The two products partition the neighbours of $v$, and which product a neighbour falls into is decided by its distance from the centre relative to the radius. In the edge centred ball a node at distance below $d$ has all of its neighbours inside $B$ and the second product is empty; the boundary sits on the distance-$d$ nodes, each keeping its inward edge resolved and closing its outward edges with messages.

Equations \eqref{eq:pv} and \eqref{eq:TB} compose in that order. Given a
configuration $\xB$, first evaluate the number $p_v(\xB)$ at every node, then hand each number to its own copy of the kernel \eqref{eq:Kkernel} and multiply. All coupling between nodes sits in the arguments of the kernels and none of it in the product: the nodes move independently, but under probabilities that the whole configuration has set. A mean-field closure differs from \eqref{eq:TB} only in replacing the argument $p_v(\xB)$ by a constant, which is what severs the coupling.

The kernel then makes the product sparse. An infectious node has a single
successor, $x_v'=x_v-1$, and contributes a factor of one; only a susceptible node branches, to $x_v'=\tau$ with probability $p_v(\xB)$ and to $x_v'=0$ with probability $1-p_v(\xB)$. Writing $S(\xB)=\{v\in B: x_v=0\}$ for the susceptible set and $A\subseteq S(\xB)$ for the subset infected on this update, the successor
$\xB'$ is fixed by $A$ alone, $x_v'=\tau$ on $A$, $x_v'=0$ on
$S(\xB)\setminus A$, and $x_v'=x_v-1$ elsewhere, and
\begin{equation}\label{eq:TBexplicit}
  T_B[\xB\to\xB']=\prod_{v\in A}p_v(\xB)\!\!\prod_{v\in S(\xB)\setminus A}\!\!\big(1-p_v(\xB)\big).
\end{equation}
A configuration therefore has $2^{|S(\xB)|}$ successors out of the
$(\tau+1)^{|B|}$ available, and \eqref{eq:TBexplicit} is what one actually builds: every reinfection probability appearing in it is a case of \eqref{eq:pv}.

\subsection{The stationary distribution of the ball}\label{sec:Pi}

The chain defined by \eqref{eq:TB} lives on the $(\tau+1)^{|B|}$ age
configurations of the ball, and its stationary distribution $\Theta$ satisfies
the same balance condition as a single node, now with one equation per
configuration:
\begin{equation}\label{eq:Pibalance}
  \Theta(\xB')=\sum_{\xB}\Theta(\xB)\,T_B[\xB\to\xB'],
\end{equation}
the weight arriving on $\xB'$ from every configuration that can reach it
equalling the weight it already carried.

The stationary law $\Theta$ is a $t\to\infty$ limit, and its temporal content sits in two places. First, the age $x_v$ is a memory, recording how long ago $v$ was infected and how long it has left, so a chain over age configurations is already a chain over histories, truncated at the depth $\tau$ that the dynamics remembers. Second, the limit sums over trajectories of every duration at their correct weight: a feedback path taking a hundred updates to close is included exactly, provided it stays inside $B$. The time axis of $\mathcal{G}$ lives inside the chain, not inside the region the chain is defined on, so $\Theta$ is the exact joint law of the in-ball ages under the closed dynamics.

Uniqueness lifts, with the all-susceptible configuration $\emptyset$ playing the
part that age $0$ played for a single node. For $0<r<1$, when the boundary
supplies infection with nonzero probability, the chain can leave $\emptyset$ and
can also fail to, so $\emptyset$ carries a self-loop and the chain is aperiodic;
and $\emptyset$ communicates with the recurrent class
reached by the closed ball, since every configuration reaches it once the
infectious ages run down. The chain is then irreducible on that class and
aperiodic, $\Theta$ is unique, and power iteration converges to it. The
disease-free boundary $\sigma=0$ is the singular limiting case used for the
threshold calculation, where $\emptyset$ is absorbing.

Factorisation does not lift, and that is the point of the construction. The
transfer operator \eqref{eq:TB} is a product over nodes, because one update acts
on each node independently once the configuration is fixed. Its stationary distribution is not: were $\Theta$ to equal $\prod_v\pi_v$, each factor would be the single-node law \eqref{eq:renewal} under some mean reinfection probability, every correlation among the ages in $B$ would vanish, and the scheme would be a mean-field closure. The departure of $\Theta$ from a product is exactly what the ball exists to capture: the echo on the central edge, the common-driver correlations of Sec.~\ref{sec:cycles}, and every higher-order correlation among its nodes. Each update acts independently on the nodes, but a stationary law is the accumulation of arbitrarily many updates, and correlation builds across them. This is how a finite region resolves correlations of unbounded temporal extent.

\subsection{Self-consistency}\label{sec:selfconsistency}

The stationary distribution is the leading left eigenvector of $T_B$, found by power iteration. Reading the conditional infectiousness \eqref{eq:msg} of one endpoint given the other is susceptible directly off $\Theta$ we find,
\begin{align}\label{eq:msgread}
  \sigma'_{i\leftarrow j}
  &=\Pball(x_j\ge1\mid x_i=0)\nonumber\\
  &=\Biggl(
     \sum\limits_{\xB:\,x_i=0,\,x_j\ge1}\Theta(\xB)
     \Biggr)
     \Bigg/
     \Biggl(
     \sum\limits_{\xB:\,x_i=0}\Theta(\xB)
     \Biggr).
\end{align}
This is a ratio of sums over configurations of the whole ball. Both endpoints are
retained in $\Theta$, so \eqref{eq:msgread} keeps the influence of $i$ on $j$
that the echo consists of, and it is through this ratio alone that the
correlations resolved inside the ball leave it.

Equation \eqref{eq:msgread} closes the model. The messages entering on the
right, through the $p_v$ of \eqref{eq:pv}, sit on the edges leaving $B_d(i,j)$,
while the message produced on the left sits on the central edge; it is therefore
one component of the system \eqref{eq:Fmap}, whose simultaneous fixed point over
all $2E$ directed edges is the model at depth $d$. One solves by iteration:
choose trial values, build the $p_v$, solve each ball, read off the updated
messages, and repeat to convergence. Algorithm 1 states the procedure.

\subsection{The prevalence}\label{sec:prevalence}

The edge calculation of Secs.~\ref{sec:transfer}--\ref{sec:selfconsistency} produces messages, not observables: solving it to convergence yields the
$\sigma_{i\leftarrow j}$ and nothing else, since the joint law $\Theta$ it computes is that of a ball built around an edge, in which no node appears in its
own right. The messages are nonetheless required to calculate the observable of interest, the prevalence of the endemic equilibrium.

To find the prevalence, we take the converged messages as given and root the
construction at a node instead. Let $\mathcal{N}_d(v)$ be the node-rooted
neighbourhood consisting of $v$, the nodes within graph distance $d$ of $v$, and
the edges among them, as in the Cantwell-Newman construction. This is a
different rooted object from the edge ball $B_d(i,j)$; the common index $d$
labels the corresponding rung of the hierarchy in the two constructions. We
build $\mathcal{N}_d(v)$, treating all $k_v$ incident edges symmetrically rather
than distinguishing a central pair, and close its boundary with those same
messages through \eqref{eq:pv}. The transfer operator is assembled exactly as in
\eqref{eq:TB} and its stationary distribution $\Theta_v$ found in the same way,
whereupon
\begin{equation}\label{eq:rhov}
  \rho_v=\sum_{\xB:\,x_v\ge1}\Theta_v(\xB),
\end{equation}
and the prevalence of the network with $N$ nodes is simply the average $\rho=N^{-1}\sum_v\rho_v$. This is a single evaluation rather than a second self-consistency problem: the messages are already fixed by
\eqref{eq:msgread}. Every correlation the node-rooted ball resolves is therefore inherited from the
edge calculation.

\subsection{What the ball resums}\label{sec:resum}

The transfer operator \eqref{eq:TB} treats every transmission between nodes of
$B$ exactly, through the indicators of \eqref{eq:pv}, so its stationary
distribution weights each trajectory of the in-ball nodes by its true
probability. The edges leaving $B$ are treated differently. To say precisely
what is exact and what is not, we write both the full process and the closed
ball as laws on \emph{histories} rather than as one-update operators, so that
the substitution the closure makes appears as a single term.

Fix a window of $T$ updates. Write $\xB(t)=\big(x_v(t)\big)_v$ for the age
configuration of the whole network at update $t$, the index running over every
node, and $\xB_B(t)$ for its restriction to $B$; the configuration $\xB$ of
Sec.~\ref{sec:transfer} is $\xB_B(t)$ with the time argument suppressed. Let
$\vec E$ denote the set of $2E$ \emph{ordered} edges, one for each direction of
each edge of the network, and write
\begin{equation}\label{eq:edgesets}
\begin{split}
  \vec E_B&=\big\{(u\to v)\in\vec E:\ u,v\in B\big\},\\
  \partial B&=\big\{(w\to v):\ v\in B,\ w\notin B,\ w\sim v\big\}
\end{split}
\end{equation}
for its in-ball part and for the background edges of Sec.~\ref{sec:hazard} taken with their inward orientation. The Bernoulli transmission variables $\eta_{u\to v}(t)$ of Sec.~\ref{sec:cycles} live on every ordered edge of the network, in-ball and background alike; write $\eta(t)$ for their collection at update $t$. A \emph{history} is one realisation of the
process,
\begin{equation}\label{eq:history}
  \omega=\big\{\xB(t)\big\}_{t=0}^{T}\cup\big\{\eta(t)\big\}_{t=0}^{T-1},
\end{equation}
of which $\omega_B$ denotes the in-ball part, namely the configurations
$\xB_B(0),\dots,\xB_B(T)$ together with the transmission variables carried by
$\vec E_B$, and $\omega_{\bar B}$ denotes everything else, $\bar B$ being the
set of nodes outside $B$.

A realisation of the transmission field becomes a trajectory once the arrival
indicator $\chi_v(t\mid x)$ of \eqref{eq:chi1} is read off it and the update
rules of Sec.~\ref{sec:model} are imposed as a Kronecker factor,
\begin{equation}\label{eq:Cv}
  \mathcal{U}_v=
  \begin{cases}
    \delta_{x_v(t+1),\,\tau\chi_v(t\mid x)}, & x_v(t)=0,\\[2pt]
    \delta_{x_v(t+1),\,x_v(t)-1},           & x_v(t)\ge1,
  \end{cases}
\end{equation}
where $\mathcal{U}_v$ abbreviates $\mathcal{U}_v\big(x_v(t+1)\mid\xB(t),\eta(t)\big)$.
Since $\tau\chi_v(t\mid x)$ takes only the values $0$ and $\tau$, the first line sets
a susceptible node to age $\tau$ exactly when it receives a transmission and
leaves it at $0$ otherwise, and the second ages an infectious node by one. The
factor vanishes on any history violating the update rules, so multiplying by it
retains the allowed histories and discards the rest.

Letting $\mu_0$ be the law of the initial configuration $\xB(0)$, and writing
\begin{equation}\label{eq:bern}
  \mathrm{Be}(b\mid q)=q^{\,b}\,(1-q)^{1-b},\qquad b\in\{0,1\},
\end{equation}
for the probability that a Bernoulli variable of parameter $q$ takes the value
$b$, written in the argument order of the kernel \eqref{eq:Kkernel}, the
probability of a history is
\begin{equation}\label{eq:fullhistory}
\begin{split}
\Prob_{\mathrm{full}}(\omega)
=\ &\mu_0\big(\xB(0)\big)\\
&\times\prod_{t=0}^{T-1}\Bigg[\prod_{(u\to v)\in\vec E}\!\!
\mathrm{Be}\big(\eta_{u\to v}(t)\mid r\big)\\
&\times\prod_{v}\mathcal{U}_v\Bigg],
\end{split}
\end{equation}
and the exact law of the in-ball history is the marginal
$\Prob_{\mathrm{full}}(\omega_B)=\sum_{\omega_{\bar B}}\Prob_{\mathrm{full}}(\omega)$, the sum
running over the exterior configurations and over every transmission variable
not carried by $\vec E_B$.

That marginal is generally not Markov in $\xB_B(t)$ alone: the exterior remembers when the ball last infected it and can return that infection many updates later, so the law of the next in-ball configuration depends on the in-ball history and not only on the current configuration. On a finite closed network it is also eventually absorbing, since with no external source the epidemic dies out and the $T\to\infty$ limit of \eqref{eq:fullhistory} is the
empty configuration. The endemic law obtained below is a genuine stationary law rather than a quasi-stationary one precisely because its boundary drive is exogenous.

The closed ball restores a Markov description by replacing the true conditional
law of the background transmissions by an independent product. For every background
edge $(w\to v)\in\partial B$ and every update, introduce
\begin{equation}\label{eq:zeta}
  \zeta_{w\to v}(t)\sim\mathrm{Bernoulli}\big(r\,\sigma_{v\leftarrow w}\big),
\end{equation}
independently over background edges and over updates, and extend
\eqref{eq:chi1} to the in-ball arrival indicator
\begin{equation}\label{eq:chiB}
\begin{split}
  \chi^B_v(t)=1-&\prod_{\substack{u\in B\\ u\sim v}}
  \big(1-\eta_{u\to v}(t)\,\one[x_u(t)\ge1]\big)\\
  &\times\prod_{\substack{w\notin B\\ w\sim v}}\big(1-\zeta_{w\to v}(t)\big),
\end{split}
\end{equation}
which reads the neighbours of $v$ inside the ball from the configuration and
those outside it from $\zeta$. Writing $\mathcal{U}^B_v$ for the factor
\eqref{eq:Cv} with $\chi_v(t\mid x)$ replaced by $\chi^B_v(t)$, and $\mu_B$ for the law of
the initial in-ball configuration, the closed-ball history law is
\begin{equation}\label{eq:ballhistory}
\begin{split}
\Prob_B(\omega_B\mid\sigma)
=\ &\mu_B\big(\xB_B(0)\big)\\
&\times\prod_{t=0}^{T-1}\Bigg[\prod_{(u\to v)\in\vec E_B}\!\!
\mathrm{Be}\big(\eta_{u\to v}(t)\mid r\big)\\
&\times\!\!\prod_{(w\to v)\in\partial B}\!\!
\mathrm{Be}\big(\zeta_{w\to v}(t)\mid r\sigma_{v\leftarrow w}\big)\\
&\times\prod_{v\in B}\mathcal{U}^B_v\Bigg],
\end{split}
\end{equation}
where $\sigma$ in the conditioning stands for the whole collection
$\{\sigma_{v\leftarrow w}\}$ on $\partial B$. The transmission variables
directed out of the ball do not appear: they act only on the exterior, which the
closure no longer represents.

Summing the Bernoulli variables out of \eqref{eq:ballhistory} returns the
transfer operator \eqref{eq:TB} with the reinfection probabilities
\eqref{eq:pv}. The step is the conditional independence of
Sec.~\ref{sec:assemble} written in the language of histories: each transmission
variable carries a directed index and therefore enters the update of its head
node alone, so given $\xB_B(t)$ the indicators $\{\chi^B_v(t)\}_{v\in B}$ are
independent, with $\E\big[\chi^B_v(t)\mid\xB_B(t)\big]=p_v\big(\xB_B(t)\big)$. The initial
law $\mu_B$ washes out under the ergodicity of Sec.~\ref{sec:Pi}, so that
$\Theta$ is the $T\to\infty$ law of $\xB_B(T)$ under \eqref{eq:ballhistory} and
does not depend on it.

What the closure replaces is not the transmission variables on $\partial B$ but the
\emph{deliveries} they make, since a background edge transmits only
when its tail is infectious. Writing
\begin{equation}\label{eq:xi}
  \xi_{w\to v}(t)=\eta_{w\to v}(t)\,\one[x_w(t)\ge1],
  \qquad (w\to v)\in\partial B,
\end{equation}
for these deliveries, whose mean given a susceptible $v$ is $r\,\sigma_{v\leftarrow w}$
by \eqref{eq:msg}, the approximation made by the ball is the single substitution
\begin{equation}\label{eq:boundaryswap}
  \Prob_{\mathrm{full}}\big(\xi\mid\omega_B\big)
  \ \longrightarrow\
  \prod_{t=0}^{T-1}\prod_{(w\to v)\in\partial B}\!\!
  \mathrm{Be}\big(\zeta_{w\to v}(t)\mid r\sigma_{v\leftarrow w}\big),
\end{equation}
and it touches nothing else: the in-ball factors of \eqref{eq:ballhistory} are
those of \eqref{eq:fullhistory} unchanged. Two properties of the discarded term
say what the substitution costs. On a locally tree-like graph $B$ separates the
subtrees lying behind distinct background edges, so conditioning on the full
in-ball history leaves them independent and nothing is lost by factorising over
$\partial B$; the whole error is carried by the single-edge law. That law is not
an independent Bernoulli sequence: $\xi_{w\to v}$ can transmit only while $w$ is
infectious, so a real neighbour delivers its attempts in a burst over the $\tau$
consecutive updates of an infectious period, and the times at which those bursts
begin depend on when the ball last fed the subtree behind it, whereas $\zeta$
arrives independently on every update at the matched rate. A sharper closure
would have to replace the Bernoulli variables on each background edge by a renewal
or age process, an enlargement in time rather than in space, which is the
direction taken by Ref.~\cite{385j-2f29} and which we return to in
Sec.~\ref{sec:discussion}. The same term shows that the conditioning in
\eqref{eq:msg} is forced rather than cosmetic: $\zeta_{w\to v}$ enters the
dynamics only through $\chi^B_v$, and $\chi^B_v$ acts only on updates for which
$v$ is susceptible, so the rate the closure must match is the conditional one.

It follows that the depth-$d$ ball treats exactly those space--time cycles all
of whose transmission edges project onto edges of $B_d$, while cycles carrying
at least one transmission edge outside $B_d$ are not discarded but retained
only through the mean rate that the substitution \eqref{eq:boundaryswap}
supplies. In the convention \eqref{eq:dist} the two families are separated by reach, a cycle of reach $L$ lying inside $B_d$ if and only if $L\le d$, so the hierarchy is graded by spatial reach much as the loopy-network construction is graded by primitive cycle length~\cite{Cantwell_Newman_2019}. Concretely, the echo lives on every edge, not only the central $(i,j)$: at $d=0$ it is resolved there and approximated on each of the remaining edges of the pair; at $d=1$ those are resolved and the approximation retreats by a shell. Unlike that construction it does not
terminate: a returning walk may travel out to any distance and come back, so
cycles of unbounded reach exist in $\mathcal{G}$ for any tree and no finite
$d$ contains them all.

At small radius the operator can be built explicitly and its stationary distribution found by power iteration, the sparsity of \eqref{eq:TBexplicit} making this inexpensive. For larger $d$ the state space grows beyond direct enumeration and $\Theta$ is obtained instead by simulating the ball. It is initialised, driven for a burn-in period, and the occupancy of the central-edge states accumulated over a long run, with each boundary node reinfected according to its own message on every update, see Algorithm 2. This is a Monte Carlo evaluation of a single application of $\Fmap_d$, not of the epidemic on a network, and the outer self-consistency loop is unchanged.

\subsection{Reduction on the configuration model}\label{sec:reduction}

Equations \eqref{eq:pv}--\eqref{eq:msgread} define the scheme on a given network: $2E$ coupled equations, one per directed edge, solved by iteration. On a configuration-model graph an annealed reduction collapses them to one scalar expression by averaging over the cavity ensemble; this reduced form is the one used for the regular-graph experiments of Sec.~\ref{sec:results}, where at fixed degree it coincides with the full per-edge system, see Fig.~\ref{fig:prevalence}.

With no degree correlations every directed edge looks out onto a cavity subtree
drawn from the same distribution, so the messages are identical in distribution.
In the annealed reduction we replace them by a single scalar $\sigma$. The boundary product
\eqref{eq:bgsurvival} then depends on the boundary node only through the number
of edges it has leaving the ball, and for a node of degree $k_v$ with $c_v$
neighbours inside $B$ it becomes $(1-r\sigma)^{k_v-c_v}$. A node reached along an
edge has its degree drawn from the excess distribution generated by $G_1$, so
this factor may be averaged. Writing $\Phi(z)=[1-r\sigma(1-z)]^{\ell}$ for the
generating function of the number of transmitting background edges of a node with
$\ell$ of them,
\begin{align}\label{eq:avgphi}
  \big\langle\Phi(z)\big\rangle
  =&\big\langle[1-r\sigma(1-z)]^{\ell}\big\rangle_{\ell\sim G_1}\nonumber\\
  =&G_1\big(1-r\sigma(1-z)\big),
\end{align}
using $\langle y^{\ell}\rangle_{\ell\sim G_1}=G_1(y)$. A susceptible node escapes
reinfection from outside the ball if and only if no background edge transmits,
with probability $\langle\Phi(0)\rangle$, so every boundary node closes with the
same \emph{background infection probability}
\begin{equation}\label{eq:hazard}
  h(\sigma)=1-G_1(1-r\sigma),
\end{equation}
the complement of that escape. The kernel \eqref{eq:Kkernel} is linear in $p$, so
using the averaged value in place of the realised one is exact at the level of a
single boundary node. It is not exact for the ball as a whole: $\Theta$ depends
nonlinearly on $T_B$, so averaging the operator is not the same as averaging its
stationary distribution. The two coincide when every node has the same degree. Retaining the full set
$\{\sigma_{i\leftarrow j}\}$ of \eqref{eq:Fmap} avoids the issue entirely, at the
cost of one ball per directed edge; a heterogeneous treatment in the spirit of
Ref.~\cite{PhysRevE.108.034310} is a natural middle course and we leave it to
future work.
Under this reduction the system \eqref{eq:Fmap} collapses to the single scalar equation
\begin{equation}\label{eq:fixedpoint}
  \sigma'=\Fmap_d(\sigma),\qquad \sigma_\ast=\Fmap_d(\sigma_\ast),
\end{equation}
with $\Fmap_d:[0,1]\to[0,1]$ fixed by $d$, $r$, $\tau$ and $G_1$. The prevalence collapses too. Two nodes of the same degree are now statistically identical. Defining the
degree-resolved prevalence by collecting all terms that have the same degree \eqref{eq:rhov},
\begin{equation}\label{eq:rhok}
  \rho(k)=\frac{\sum_v \delta_{k_v,k}\,\rho_v}{\sum_v \delta_{k_v,k}},
  \qquad \sum_v \delta_{k_v,k}=N p_k ,
\end{equation}
the network prevalence follows from $\sum_k\delta_{k_v,k}=1$ as
\begin{equation}\label{eq:rhoCM}
  \rho=\frac1N\sum_v \rho_v=\sum_k p_k\,\rho(k).
\end{equation}

\subsection{The endemic threshold}\label{sec:threshold}

The disease-free message vector is a fixed point of \eqref{eq:Fmap} at
every depth: with nothing arriving on the boundary the ball settles into the
empty configuration and emits nothing. A monotone-coupling argument would
settle the stability of the full nonlinear map, but the fixed-period dynamics
is not monotone in the transmission field, as the example of
Sec.~\ref{sec:cycles} shows, and we do not argue that way. The threshold needs
only the linearisation at this fixed point, which is a non-negative
$2E \times 2E$ Jacobian $\mathcal{J}_d(r)$; loss of stability is therefore governed by its
Perron root $\lambda_d(r)$, and the threshold is fixed by
$\lambda_d\big(r_c^{(d)}\big) = 1$.

Under the configuration-model reduction of Sec.~\ref{sec:reduction} the same
linearisation collapses to a scalar slope. We write this reduced growth factor
as
\begin{equation}
  \Lambda_d \;\equiv\; \Fmap_d'(0)
  \;=\; \left. \frac{\mathrm{d}\sigma'}{\mathrm{d}\sigma} \right|_{\sigma = 0} ,
  \label{eq:Lambda}
\end{equation}
with the reduced threshold given by $\Lambda_d\big(r_c^{(d)}\big) = 1$; on a
regular graph, where the reduction is exact, the Perron vector of $\mathcal{J}_d$ is
uniform and $\lambda_d = \Lambda_d$. Above the threshold an endemic fixed point
exists; uniqueness would require concavity of $\Fmap_d$ on $[0,1]$, which
we observe numerically at every depth examined but do not establish.

Equation \eqref{eq:Lambda} is evaluated without depth-specific algebra. Because
$\Fmap_d(0) = 0$ exactly, the slope is the directional derivative of the
map at the origin, and this follows in closed form from the linear response of
the ball at $\sigma = 0$. Appendix~\ref{app:algorithm} gives that construction,
together with the unbiased estimator that replaces enumeration once the state
space of $B_d$ outgrows it, and $r_c^{(d)}$ then follows by bisection on $r$.

It is worth recording the value obtained when the echo is deleted, since it
anchors what the hierarchy corrects. Forbidding the return transmission
decouples the two endpoints of the central edge, each reducing to the driven
node of Sec.~\ref{sec:atom}; linearising \eqref{eq:renewal} with
$p = h(\sigma)$ of \eqref{eq:hazard} gives
\begin{equation}
  \Lambda_{\mathrm{nb}} \;=\; \tau \, G_1'(1) \, r ,
  \qquad
  r_c^{\mathrm{nb}} \;=\; \frac{1}{\tau G_1'(1)} ,
  \label{eq:nbthreshold}
\end{equation}
the threshold of recurrent dynamic message passing \cite{Shrestha_Scarpino_Moore_2015}, which forbids
that transmission by construction.

The bottom rung of the hierarchy admits a closed form, and may be compared with
\eqref{eq:nbthreshold} directly. At $d = 0$ the ball is the pair $\{i,j\}$, and
at $\sigma = 0$ it receives no drive from outside, so the trace summed by
\eqref{eq:slopemc} is that of an infection seeded at one endpoint and passed
back and forth across the central edge until it fails to be returned. Writing
$V(e_v)$ for the expectation in \eqref{eq:slopemc} conditioned on the
seed $e_v$, first-step analysis over the $(\tau+1)^2$ configurations of
the pair gives
\begin{equation}
  V(e_i) + V(e_j) \;=\; \frac{\tau}{1 - r} ,
  \label{eq:pairtrace}
\end{equation}
for every $\tau \geq 1$ and independently of the degree distribution, which
enters only through the source. Under the reduction each endpoint carries an
excess degree averaged by $G_1$, so that
$s(e_i) = s(e_j) = r \, G_1'(1)$, the two seeds are drawn with
equal probability, and $|s| = 2 r \, G_1'(1)$. Equation \eqref{eq:slopemc} then
returns
\begin{equation}
  \Lambda_0 \;=\; \frac{\tau \, G_1'(1) \, r}{1 - r} ,
  \qquad
  r_c^{(0)} \;=\; \frac{1}{1 + \tau G_1'(1)} .
  \label{eq:depth0}
\end{equation}
Comparing \eqref{eq:depth0} with \eqref{eq:nbthreshold},
\begin{equation}
  \Lambda_0 \;=\; \frac{\Lambda_{\mathrm{nb}}}{1 - r}
  \;=\; \Lambda_{\mathrm{nb}} \sum_{n \geq 0} r^{\,n} ,
  \label{eq:resummation}
\end{equation}
so the depth-$0$ ball multiplies the non-backtracking growth factor by a
geometric series in which $r^{\,n}$ carries the histories that recross the
central edge $n$ further times. The echo of Sec.~\ref{sec:cycles} is thereby
resummed to all orders on that edge, and each rung above $d = 0$ extends the
same resummation to cycles of greater reach; \eqref{eq:resummation} is the
precise form of the correction discussed in Sec.~\ref{sec:discussion}, where the
leading backtracking contribution identified in Ref.~\cite{Castellano_PastorSatorras_2018} is the
$n = 1$ term. For $\tau = 2$ on the $3$-regular graph,
$r_c^{\mathrm{nb}} = 1/4$ against $r_c^{(0)} = 1/5$, and the deeper rungs of
Sec.~\ref{sec:results} move back toward the simulated onset from below.

\section{Results}\label{sec:results}

\begin{figure}[!htb]
\centering
\includegraphics[width=\columnwidth]{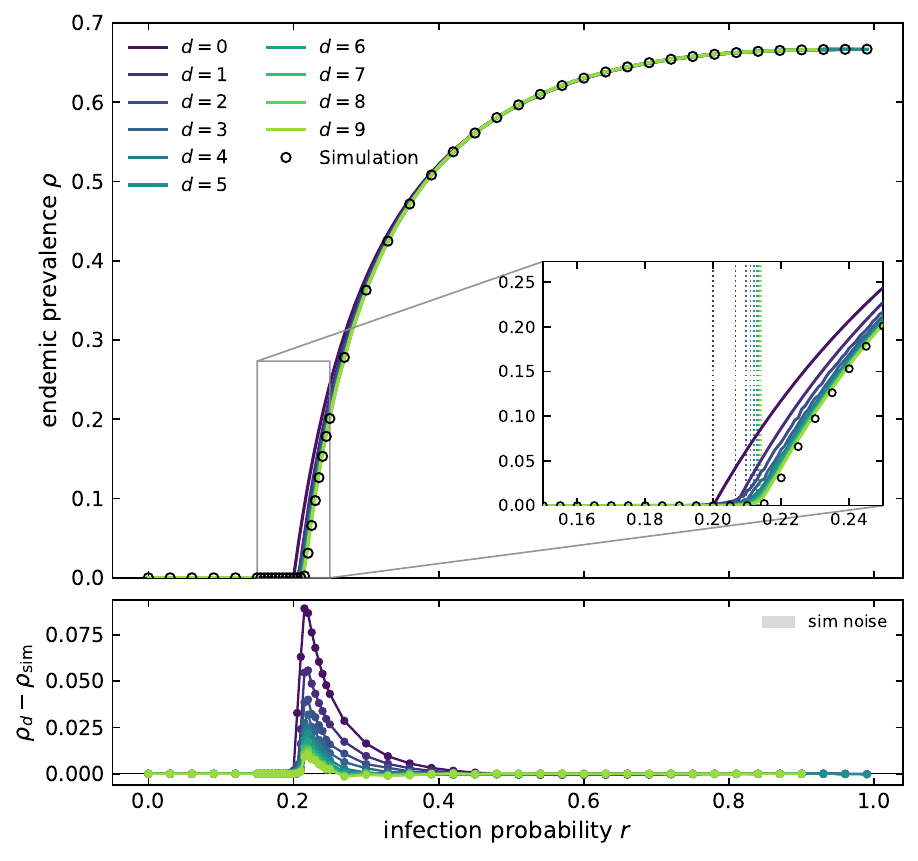}
\caption{(top) The prevalence of infected nodes at the endemic equilibrium as a function of the infection probability $r$, on a random $3$-regular graph with $\tau=2$ and $N=150{,}000$ nodes. Curves are the members of the hierarchy; points are direct simulation. The inset shows the region around the critical point, with the thresholds of each rung marked and a finer gridding. (bottom) The residual between simulation and theory, with the shaded band showing the simulation noise floor being smaller than the plot line.}
\label{fig:prevalence}
\end{figure}

Figure~\ref{fig:prevalence} shows the stationary prevalence of Eq.~\ref{eq:rhoCM} for $\tau=2$ on a
random $3$-regular graph with $N=150{,}000$ nodes at increasing $d$. The deeper rungs match direct Monte Carlo simulation scatter points across the entire endemic range. The inset shows the detail of the threshold region with the numerically-obtained endemic threshold values for each $d$ plotted as vertical lines. The theoretical endemic thresholds, found by numerical solution of Eq.~\eqref{eq:Lambda}, increase with $d$ over the depths shown and move toward the onset region of the simulated endemic branch. The model--simulation residual, $\rho_d-\rho_{\text{sim}}$, is plotted in the lower panel. The magnitude of the residual is not constant over the range of infection probabilities, taking its largest value at the threshold and falling monotonically with infection probability $r$. It also falls monotonically with increasing $d$. The simulation noise is also plotted on the bottom chart and is smaller than the lines.

\begin{figure}
 \includegraphics[width=\columnwidth]{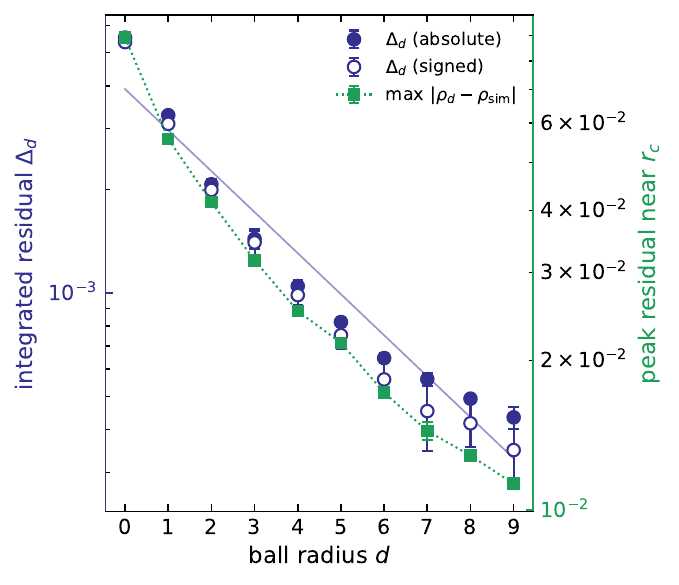} 
\caption{Convergence of the hierarchy for the experiment in Fig.~\ref{fig:prevalence}. The integrated residual $\Delta_d$ of Eq.~\eqref{eq:Delta} against the radius $d$ of the resolved ball. Filled symbols, the absolute integral; open symbols, the signed integral, which is insensitive to the sampling noise of the rungs solved by simulation. The vertical scale is logarithmic, on which a geometric decay in $d$ is a straight line.}
\label{fig:convergence}          
\end{figure}

The residual of Fig.~\ref{fig:prevalence} can be collapsed to a single number for each rung. Following~\cite{pgxg-mhjr}, we define
  \begin{equation}
    \Delta_d\equiv\int_0^1\big|\rho_d(r)-\rho_{\rm sim}(r)\big|\,\mathrm{d}r,
    \label{eq:Delta}
  \end{equation}
the area between the depth-$d$ prevalence and the simulated one across the endemic range. Figure~\ref{fig:convergence} shows $\Delta_d$ against $d$. It falls monotonically, and by a roughly constant factor per shell over the depths examined, consistent with a geometric decay in the radius of the resolved ball: each additional shell of exactly treated space--time cycles removes an approximately fixed fraction of the remaining error rather than a fixed amount. The integrand is concentrated just above the threshold.

We plot the signed error alongside the absolute one because the sampling error of the deeper rungs enters the two differently. It enters $\int(\rho_d-\rho_{\rm sim})\,\mathrm{d}r$ linearly, so it cancels between points and averages down with the sample count; it enters \eqref{eq:Delta} through a modulus, which cannot cancel. We observe that the two integrals agree closely at every depth and fall monotonically together. This indicates both that $\rho_d-\rho_{\rm sim}$ is of one sign across the endemic range and that the rectified noise is too small to disturb the ordering of Fig.~\ref{fig:convergence}.

The single sign of this error, and the approach of the threshold from below, we attribute to the boundary substitution of Sec.~\ref{sec:resum}. A real neighbour is infectious in bursts of $\tau$ consecutive updates, so that, conditioned on $v$ being susceptible, its transmission attempts are positively autocorrelated in time: once one succeeds and infects $v$, the remaining attempts of that burst fall on an already-infectious node and are wasted. The Bernoulli surrogate $\zeta$ carries the same conditional mean rate but transmits independently, so its attempts are spread across more distinct susceptible spells and a smaller fraction are wasted within one infectious window. The surrogate is therefore the more efficient transmitter, so the closure overestimates the prevalence and underestimates the threshold at every depth; enlarging the ball pushes this bias outward onto cycles of greater reach but does not remove it. This is consistent with $\rho_d-\rho_{\rm sim}$ keeping a single sign and the depth-$d$ threshold approaching the simulated onset from below.

\begin{figure}[!htb]
\centering
\includegraphics[width=\columnwidth]{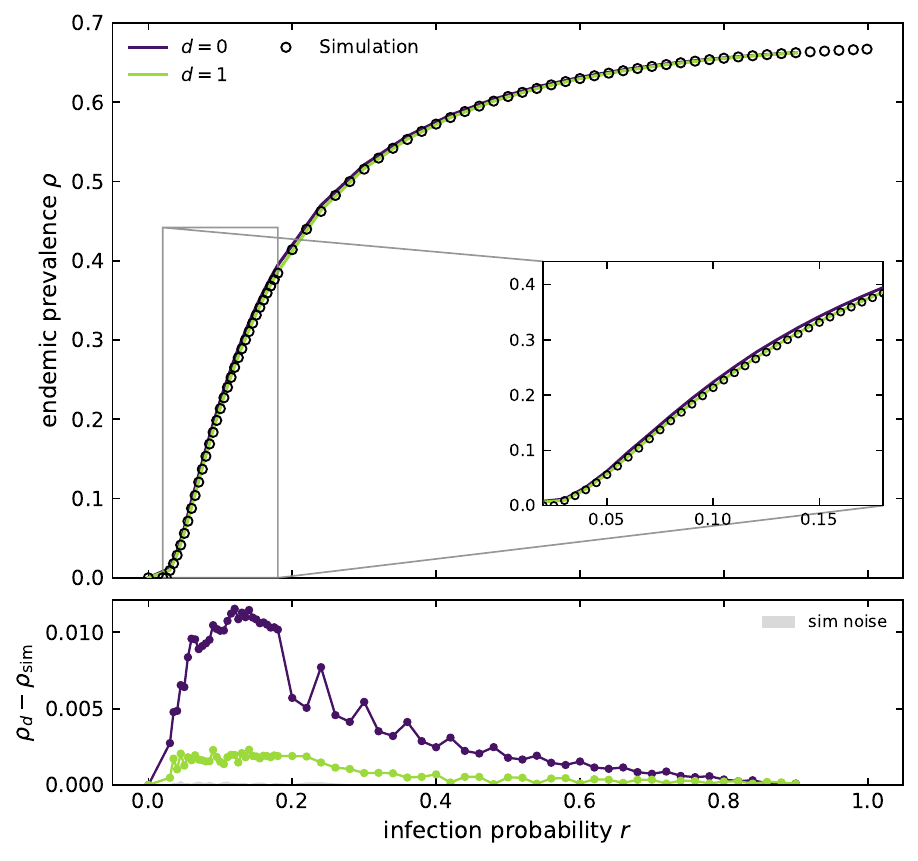}
\caption{(top) The prevalence of the endemic equilibrium of largest component of a coauthorship network of 13,861 scientists \cite{Newman_2001}. The inset shows a finer gridding around the critical region. (bottom) The residual between simulation and the model.}
\label{fig:condmat_peredge}
\end{figure}

In Fig.~\ref{fig:condmat_peredge} we plot the prevalence of the endemic equilibrium for the largest connected component of a 13,861 scientist coauthorship network Ref.~\cite{Newman_2001}; finding excellent agreement with Monte Carlo simulation. These curves are computed with the full per-edge system of \eqref{eq:Fmap} rather than the configuration-model reduction of Sec.~\ref{sec:reduction}, whose degree-independent cavity assumption fails on a coauthorship network due to the high density of short range loops. This empirical network is known to have a high density of short loops and is therefore a challenging network to study, even for ordinary SIR or bond percolation. The agreement is nonetheless excellent because the ball $B_d$ contains every edge among its nodes: any short spatial loop that falls inside the ball is absorbed and treated exactly, as in the loopy-network construction of Ref.~\cite{Cantwell_Newman_2019}, so that only loops straddling the boundary are approximated. The magnitude of the residual is much smaller at the critical point than the residual for $d=0,1$ of the lattice experiment. Compared with the lower panel of Fig.~\ref{fig:prevalence}, the residual also decays more broadly with $r$.

\section{Discussion}\label{sec:discussion}

We have shown that a recurrent disease on a locally tree-like network generates
cycles in space--time that are closed by the persistence of a node's own state rather
than by any spatial loop. These cycles obstruct message passing in exactly the way that spatial loops obstruct it for ordinary percolation. The obstruction is the backtracking walk that the cavity construction is designed to forbid, and the remedy is to restore it in a controlled way: solve the dynamics exactly inside a ball about an edge, close the boundary with a single conditional message, and let the radius grow. The construction delivers the endemic threshold
at every depth.

Three other treatments of the same correlation deserve comment. Shrestha, Scarpino and Moore~\cite{Shrestha_Scarpino_Moore_2015} identified it as the echo chamber effect and suppressed it, forbidding causal signals from backtracking to their immediate source. The depth-$0$ correction is the difference between suppressing that correlation and resolving it. Castellano and Pastor-Satorras~\cite{Castellano_PastorSatorras_2018} isolate the role of
backtracking paths in recurrent-state spreading and quantify the error they induce on a non-backtracking closure; their diagnosis is the one the present construction is built to repair, and the depth-$0$ correction may be read as the leading contribution they identify, resummed on the central edge.

Cantwell and Moore~\cite{385j-2f29} improve the pair approximation for
continuous-time SIS by augmenting the susceptible state with a chain of $K$ states acting as an approximate clock for how long a node has been susceptible, so that a pair can carry the correlation between that elapsed time and its neighbour's state. Both schemes recognise that the pair fails for SIS because the time since a node's last infection is informative about its neighbours, and both repair it by enlarging the state space; they differ in the direction of the
enlargement. Cantwell and Moore enlarge in time, refining the resolution with which a single pair remembers its own history, while the ball enlarges in space, resolving the neighbourhood through which the returning walk actually travels. The fixed infectious period plays the role in our construction that the $K$-state clock plays in theirs: in both cases it is the device that renders a node's memory finite. The two hierarchies are therefore complementary, and a scheme indexed by both radius and clock resolution is a natural object we do not pursue here. A quantitative comparison of thresholds is complicated by the underlying dynamics. Ref.~\cite{385j-2f29} treats the continuous-time contact process with exponentially distributed recovery, whereas the present model is discrete-time with a deterministic infectious period, so that the contact-process critical infection rate $\beta_c$ and our per-update threshold $r_c$ are not directly commensurable, and we do not attempt to force the comparison.

A parallel treatment of dynamics on sparse graphs comes from the cavity method itself. The dynamical cavity method takes a node's trajectory over $T$ updates as a single augmented variable and applies the static cavity method to it~\cite{J_P_L_Hatchett_2004,Mimura_2009,kanoria2011majority,Neri_2009}, which is exact at finite $T$ but costs exponentially in it; the unidirectional case is the one that closes without approximation~\cite{Lokhov2015Dynamic}, for the reason set out in Sec.~\ref{sec:cycles}. Two features of that work bear on the construction here. First, the factor graph of a synchronous pairwise dynamics fails to be locally tree-like even on a tree, since $i$ and a neighbour $j$ enter each other's update factors and so close a loop of length four on every edge; passing to an edge-dual representation, whose variables are pairs of neighbouring trajectories, removes it~\cite{PhysRevX.13.031021}. Our ball is centred on an edge for the same reason, and $B_0$ is that pair. Second, reaching the long-time limit is the common difficulty: the backtracking dynamical cavity method recovers the attractors of a deterministic dynamics by imposing a periodicity condition on a finite window~\cite{PhysRevX.13.031021}\footnote{The ``backtracking'' of Ref.~\cite{PhysRevX.13.031021} is in time, tracing
trajectories back from an attractor, and is unrelated to the backtracking walk
$i\to j\to i$ of Sec.~\ref{sec:cycles}.}, where the fixed period $\tau$ instead makes the chain on the ball finite, so that the $t\to\infty$ limit is taken inside the region rather than along the variable. As with the clock of Ref.~\cite{385j-2f29}, the enlargement in that literature is in time, while the ball enlarges in space.

The construction is not specific to SIS. Recurrent processes with waning immunity, reactivation or refractory binary-state dynamics generate the same returning walks and can be closed by the same ball-indexed hierarchy. Multiple coexisting recurrent diseases \cite{Mann_Smith_Mitchell_Dobson_2021,Mann_Smith_Mitchell_Dobson_2022}, and mixtures of progressive and recurrent dynamics, could be accommodated in the same way: the per-node state enlarges, but the structure of the ball is untouched. Infectious periods drawn from a distribution are a harder matter, since the finite memory that makes the ball tractable must then be recovered by other means.

A computational improvement would be to lower the cost of solving the stationary dynamics of the ball, which is what limits the depth attainable in practice. Where the network admits an edge-disjoint motif cover, the ball should be assembled from its constituent cliques and chordless cycles rather than enumerated node by node, in the manner of Ref.~\cite{PhysRevE.107.054303}. Also, the depth need not be uniform: following Ref.~\cite{PhysRevE.108.034310}, each edge could carry its own radius, so that depth is spent only where the returning walk is strong enough to warrant it. Also, the ball need not be symmetric about the terminal nodes of the focal edge. The outstanding theoretical question is the truncation error, whose control would require the returning excursions outside a ball to be organised into a convergent series rather than merely counted. Other techniques, including matrix-product representations or tensor compression~\cite{PhysRevE.97.010104,PhysRevLett.132.117401}, could also be used to solve the dynamics inside the ball; we leave this for future work.

\appendix

\section{Computational methods}\label{app:algorithm}

We present an algorithm to solve the message passing dynamics on a network. We note that computing the exact stationary dynamics inside the ball for large radius is computationally costly and thus provide two methods to evaluate it: power iteration and simulation. For the endemic threshold the linearisation can be taken exactly, and the resulting response either enumerated or sampled; the sampled form removes the state space from the calculation altogether and is what makes large $d$ accessible. In this section we will discuss computation techniques to keep the calculations feasible when increasing $d$.

\begin{widetext}
\noindent\rule{\textwidth}{0.8pt}\\[2pt]
\textbf{Algorithm 1.} Depth-$d$ ball message passing.\\[-6pt]
\noindent\rule{\textwidth}{0.4pt}
\begin{algorithmic}[1]
\Require network, infection probability $r$, period $\tau$, radius $d$,
         tolerance $\epsilon$
\State $\sigma_{i\leftarrow j}\gets\sigma_{\rm init}$ for every directed edge
         \Comment{any value in $(0,1]$}
\Statex
\Statex \Comment{outer loop: solve the message system \eqref{eq:Fmap}}
\Repeat
  \State $\delta\gets0$
  \ForAll{directed edges $(i,j)$}
    \State assemble $B_d(i,j)$
    \State build $p_v(\xB)$ for all $v\in B_d(i,j)$ from Eq.~\eqref{eq:pv},
           using the current messages on the edges leaving the ball
    \If{$|B_d(i,j)|$ small enough to enumerate}
      \State build $T_B$ from Eq.~\eqref{eq:TBexplicit}
      \State $\Theta\gets$ leading left eigenvector of $T_B$ by power iteration
    \Else
      \State $\Theta\gets$ occupancy histogram from simulating the ball
    \EndIf
    \State $\sigma^{\rm new}_{i\leftarrow j}\gets
            \big(\sum_{\xB:\,x_i=0,\,x_j\ge1}\Theta(\xB)\big)\big/
            \big(\sum_{\xB:\,x_i=0}\Theta(\xB)\big)$
           \Comment{Eq.~\eqref{eq:msgread}}
    \State $\delta\gets\max\{\delta,\,
            |\sigma^{\rm new}_{i\leftarrow j}-\sigma_{i\leftarrow j}|\}$
  \EndFor
  \State $\sigma\gets\sigma^{\rm new}$ on every directed edge
\Until{$\delta<\epsilon$}
\Statex
\Statex \Comment{prevalence: one pass, re-rooted at nodes, no further iteration}
\State $\rho\gets0$
\ForAll{nodes $v$}
  \State assemble the radius-$d$ ball rooted at $v$, retaining all $k_v$ edges
  \State close its boundary with the converged messages and solve for $\Theta_v$
  \State $\rho\gets\rho+\sum_{\xB:\,x_v\ge1}\Theta_v(\xB)$
           \Comment{Eq.~\eqref{eq:rhov}}
\EndFor
\State $\rho\gets\rho/N$
\State \Return $\{\sigma_{i\leftarrow j}\}$, $\rho$
\Statex
\Statex \Comment{on a configuration-model graph the messages are all equal, the}
\Statex \Comment{edge loop collapses to one ball, and the node loop to a sum}
\Statex \Comment{over degrees weighted by $p_k$; see Sec.~\ref{sec:reduction}}
\end{algorithmic}
\noindent\rule{\textwidth}{0.8pt}

\noindent\rule{\textwidth}{0.8pt}\\[2pt]
\textbf{Algorithm 2.} Stationary distribution of a ball by simulation
(line 11 of Algorithm 1).\\[-6pt]
\noindent\rule{\textwidth}{0.4pt}
\begin{algorithmic}[1]
\Require ball $B$, messages $\sigma$ on the edges leaving it, infection
         probability $r$, period $\tau$, burn-in $T_{\rm burn}$, sample count $T_{\rm s}$
\ForAll{$v\in B$}
  \State $q_v\gets\prod_{w\notin B,\,w\sim v}\!\big(1-r\,\sigma_{v\leftarrow w}\big)$
         \Comment{Eq.~\eqref{eq:bgsurvival}; fixed for the whole run}
\EndFor
\State $x_v\gets$ uniform in $\{1,\dots,\tau\}$ for all $v\in B$
       \Comment{start fully infectious: the endemic branch is approached from above}
\State zero the counters $C_{\rm s},\,C_{\rm si},\,C_{\rm s}',\,C_{\rm si}'$ and
       $C_v$ for all $v\in B$
\Statex
\For{$t=1,\dots,T_{\rm burn}+T_{\rm s}$}
  \ForAll{$v\in B$}
    \If{$x_v=0$}
      \State $n_v\gets\#\{u\in B:\,u\sim v,\;x_u\ge1\}$
             \Comment{in-ball neighbours read from the current $\xB$}
      \State $p_v\gets1-q_v\,(1-r)^{n_v}$ \Comment{Eq.~\eqref{eq:pv}}
      \State $x'_v\gets\tau$ with probability $p_v$, else $x'_v\gets0$
    \Else
      \State $x'_v\gets x_v-1$ \Comment{ageing; an infectious node has one successor}
    \EndIf
  \EndFor
  \State $\xB\gets\xB'$
         \Comment{all transmissions use the configuration at the start of the update}
  \If{$t>T_{\rm burn}$}
    \State $C_v\gets C_v+1$ for every $v$ with $x_v\ge1$
    \State $C_{\rm s}\gets C_{\rm s}+[x_i=0]$, \quad
           $C_{\rm si}\gets C_{\rm si}+[x_i=0][x_j\ge1]$
    \State $C_{\rm s}'\gets C_{\rm s}'+[x_j=0]$, \quad
           $C_{\rm si}'\gets C_{\rm si}'+[x_j=0][x_i\ge1]$
  \EndIf
\EndFor
\Statex
\State \Return $\sigma_{i\leftarrow j}=C_{\rm si}/C_{\rm s}$ and
       $\sigma_{j\leftarrow i}=C_{\rm si}'/C_{\rm s}'$ \Comment{Eq.~\eqref{eq:msgread}}
\State \Return $\rho_v=C_v/T_{\rm s}$ for a ball rooted at a node \Comment{Eq.~\eqref{eq:rhov}}
\end{algorithmic}
\noindent\rule{\textwidth}{0.8pt}

\noindent\rule{\textwidth}{0.8pt}\\[2pt]
\textbf{Algorithm 3.} Direct simulation of the network, the reference of
Sec.~\ref{sec:results}.\\[-6pt]
\noindent\rule{\textwidth}{0.4pt}
\begin{algorithmic}[1]
\Require network on $N$ nodes, infection probability $r$, period $\tau$,
         burn-in $T_{\rm burn}$, sample count $T_{\rm s}$, runs $R$, initial fraction $f_0$
\State $\rho\gets0$, \; $R_{\rm eff}\gets0$
\For{$\alpha=1,\dots,R$}
  \State $x_v\gets$ uniform in $\{1,\dots,\tau\}$ with probability $f_0$, else $0$
  \State $\Sigma\gets0$, \; $m\gets0$
  \For{$t=1,\dots,T_{\rm burn}+T_{\rm s}$}
    \ForAll{$v$}
      \If{$x_v=0$}
        \State $n_v\gets\#\{u\sim v:\,x_u\ge1\}$
        \State $x'_v\gets\tau$ with probability $1-(1-r)^{n_v}$, else $x'_v\gets0$
               \Comment{no boundary term: every neighbour is resolved}
      \Else
        \State $x'_v\gets x_v-1$
      \EndIf
    \EndFor
    \State $\xB\gets\xB'$, \quad $I\gets\#\{v:\,x_v\ge1\}$
    \If{$I=0$} \State \textbf{break} \Comment{the epidemic has died; later updates carry no information}
    \EndIf
    \If{$t>T_{\rm burn}$} \State $\Sigma\gets\Sigma+I/N$, \; $m\gets m+1$ \EndIf
  \EndFor
  \If{$m>0$} \State $\rho\gets\rho+\Sigma/m$, \; $R_{\rm eff}\gets R_{\rm eff}+1$ \Comment{accumulate only runs that survived burn-in} \EndIf
\EndFor
\State \Return $\rho/R_{\rm eff}$ and $R_{\rm eff}$, with the standard error taken over the $R_{\rm eff}$ surviving runs
\Statex
\Statex \Comment{Updates (i)--(iii) of Sec.~\ref{sec:model} are applied exactly as in}
\Statex \Comment{Algorithm 2; the two differ only in that the network has no boundary}
\Statex \Comment{to close, so no message enters. Averaging over the surviving portion}
\Statex \Comment{of a run is a quasi-stationary estimate: near the threshold a finite}
\Statex \Comment{network reaches the absorbing state at a rate that vanishes with $N$,}
\Statex \Comment{and the unconditional average would report the absorbing state rather}
\Statex \Comment{than the endemic one.}
\end{algorithmic}
\noindent\rule{\textwidth}{0.8pt}
\end{widetext}
\subsection{Sparsity of the enumerated solve}

The state space $(\tau+1)^{|B|}$ overstates the cost of the stationary solve in
two ways, and we exploit both.

The first is structural and is visible in Eq.~\eqref{eq:TBexplicit}: an infectious node has a single successor and contributes a factor of one, so only the susceptible set branches and a configuration has $2^{|S(\xB)|}$ successors. We therefore never build $T_B$ as a matrix. Each configuration is pushed onto its successors by a depth-first walk over $S(\xB)$, in which the reinfection probabilities of Eq.~\eqref{eq:pv} enter as branch weights; branches whose accumulated weight falls below a floor are cut, and the resulting distribution is renormalised.

The second is dynamical. $\Theta$ is concentrated on a small part of the state space, sharply so near the disease-free fixed point. We keep an active list of the configurations carrying weight, expand only those, and rebuild the list from the configurations reached. The cost of an iteration then tracks the support of $\Theta$ rather than the size of the ball.

\subsection{Frozen randomness}

When the ball is simulated rather than enumerated, the random stream is derived from the ball, the radius and the index of $r$, and never from the outer-iteration counter. Under that convention $\Fmap_d$ is a deterministic function of $\sigma$, its fixed point is well defined, and the $\delta$ of Algorithm~1 measures the movement of the messages rather than the movement of the sampler; sampling noise enters as a bias controlled by the sample count, which we estimate from runs at independent seeds. Cantwell and Newman~\cite{Cantwell_Newman_2019} reuse a single sequence of edge additions across successive iterations of their message passing for the same reason.

\subsection{Continuation in $r$}

We sweep $r$ downwards in the numerical continuation and start each point from
the solution of the previous one. In the cases studied this keeps the iteration
close to the endemic branch and reduces the number of outer iterations
substantially; it is a computational device rather than a monotone-coupling
proof of the nonlinear map.

\subsection{Linearisation of the message map}

The linearisation at the disease-free state is obtained in closed form. Let
$T_0$ denote the transfer operator at $\sigma=0$ and $T_1$ its directional
derivative at the origin along a direction $u$ in the messages, and let
$\Theta_1$ be the corresponding derivative of the stationary distribution.
Differentiating
$\Theta=\Theta T_B$ at $\sigma=0$, where $\Theta=\delta_\varnothing$, gives
\begin{equation}
  \Theta_1\,(I-T_0)=\delta_\varnothing T_1 .
  \label{eq:linresp}
\end{equation}
The empty configuration is absorbing under $T_0$, so $\delta_\varnothing$ is a
left fixed vector of $T_0$ and \eqref{eq:linresp} closes on its complement.
Writing $A$ for $T_0$ restricted to the non-empty configurations and $\Psi$ for
the restriction of $\Theta_1$,
\begin{equation}
  \Psi=s\,(I-A)^{-1}=\sum_{t\ge0}s\,A^{t},
  \qquad
  s(e_v)=r\!\!\sum_{\substack{w\notin B\\ w\sim v}}\!\! u_{v\leftarrow w},
  \label{eq:neumann}
\end{equation}
where $e_v$ is the configuration carrying $v$ at age $\tau$ and every other
node susceptible. At $\sigma=0$ the denominator of Eq.~\eqref{eq:msgread} is
unity, so differentiating it leaves
\begin{equation}
  \frac{\partial\sigma'_{i\leftarrow j}}{\partial u}
  =\sum_{\xB:\,x_i=0,\,x_j\ge1}\Psi(\xB),
  \label{eq:slopelin}
\end{equation}
This is the action of the unreduced Jacobian $\mathcal{J}_d(r)$ on the
perturbation $u$. Its Perron root $\lambda_d(r)$ is recovered by power
iteration over the directed edges. Under the reduction of
Sec.~\ref{sec:reduction}, the same calculation is a single evaluation and gives
the scalar $\Lambda_d$ of Eq.~\eqref{eq:Lambda}.

For $r<1$ every configuration has positive probability of transmitting nothing
for $\tau$ consecutive updates, so infection dies out inside the ball, the
spectral radius of $A$ is below $1$, and \eqref{eq:neumann} converges geometrically. Its terms are
non-negative, so the partial sums are lower bounds on $\Lambda_d$ and a
bisection step is stopped as soon as one exceeds unity.

\subsection{Sampling the response}

Equation~\eqref{eq:neumann} sums over the histories of an infection seeded at a
single node and propagated with the boundary closed, and may be sampled rather
than enumerated. Writing $|s|=\sum_v s(e_v)$, we draw a seed node with
probability $s(e_v)/|s|$, run the ball forward at zero boundary rate until the
infection dies out, and accumulate the indicator of Eq.~\eqref{eq:msgread}
along the trace, giving
\begin{equation}
  \Lambda_d=|s|\;
  \mathbb E\!\left[\sum_{t\ge0}
  \mathbf 1\big(x_i(t)=0,\;x_j(t)\ge1\big)\right].
  \label{eq:slopemc}
\end{equation}
The estimator is unbiased, the sampled quantity is of order unity, and no
distribution over configurations is ever represented: the storage is of order
$|B|$. As above, the trace seeds are derived from the ball and not from $r$, so
that $\Lambda_d(r)$ is a deterministic function of $r$ across the bisection.

For $\tau=2$ on the $3$-regular tree the depth-$5$ ball carries $|B|=126$ nodes and more than $10^{60}$ configurations; the thresholds at $d=0,\dots,5$ are
obtained together in a few minutes on a single core, and where both routes are
available the sampled and enumerated thresholds agree to $2\times10^{-4}$.

\bibliography{ref}

@article{PhysRevLett.116.258301,
  title = {Solving the Dynamic Correlation Problem of the Susceptible-Infected-Susceptible Model on Networks},
  author = {Cai, Chao-Ran and Wu, Zhi-Xi and Chen, Michael Z. Q. and Holme, Petter and Guan, Jian-Yue},
  journal = {Phys. Rev. Lett.},
  volume = {116},
  issue = {25},
  pages = {258301},
  numpages = {5},
  year = {2016},
  month = {Jun},
  publisher = {American Physical Society},
  doi = {10.1103/PhysRevLett.116.258301},
  url = {https://link.aps.org/doi/10.1103/PhysRevLett.116.258301}
}

@article{Newman_2001, title={The structure of Scientific Collaboration Networks}, volume={98}, DOI={10.1073/pnas.98.2.404}, number={2}, journal={Proceedings of the National Academy of Sciences}, author={Newman, M. E.}, year={2001}, month={Jan}, pages={404–409}}

@article{PhysRevLett.104.258701,
  title = {Epidemic Threshold for the Susceptible-Infectious-Susceptible Model on Random Networks},
  author = {Parshani, Roni and Carmi, Shai and Havlin, Shlomo},
  journal = {Phys. Rev. Lett.},
  volume = {104},
  issue = {25},
  pages = {258701},
  numpages = {4},
  year = {2010},
  month = {Jun},
  publisher = {American Physical Society},
  doi = {10.1103/PhysRevLett.104.258701},
  url = {https://link.aps.org/doi/10.1103/PhysRevLett.104.258701}
}

@book{levin2008markov,
  title     = {Markov Chains and Mixing Times},
  author    = {Levin, David A. and Peres, Yuval and Wilmer, Elizabeth L.},
  year      = {2008},
  publisher = {American Mathematical Society},
  address   = {Providence, Rhode Island},
  isbn      = {978-0-8218-4739-1}
}

@book{Norris_1997, place={Cambridge}, series={Cambridge Series in Statistical and Probabilistic Mathematics}, title={Markov Chains}, publisher={Cambridge University Press}, author={Norris, J. R.}, year={1997}, collection={Cambridge Series in Statistical and Probabilistic Mathematics}}

@article{wg82-f4lf,
  title = {Epidemic threshold and localization of the SIS model on directed complex networks},
  author = {M\"uller, Vin\'{\i}cius B. and Metz, Fernando L.},
  journal = {Phys. Rev. E},
  volume = {112},
  issue = {6},
  pages = {064303},
  numpages = {12},
  year = {2025},
  month = {Dec},
  publisher = {American Physical Society},
  doi = {10.1103/wg82-f4lf},
  url = {https://link.aps.org/doi/10.1103/wg82-f4lf}
}

@inproceedings{conf/aaai/Pearl82,
  author    = {Judea Pearl},
  title     = {Reverend {Bayes} on Inference Engines: A Distributed Hierarchical Approach},
  booktitle = {Proceedings of the National Conference on Artificial Intelligence (AAAI)},
  year      = {1982},
  pages     = {133--136},
  publisher = {AAAI Press},
  isbn      = {0-262-51051-0}
}

@article{10.1098/rspa.1935.0122,
    author = {Bethe, H. A.},
    title = {Statistical theory of superlattices},
    journal = {Proceedings of the Royal Society of London. A. Mathematical and Physical Sciences},
    volume = {150},
    number = {871},
    pages = {552-575},
    year = {1935},
    month = {07},
    issn = {0080-4630},
    doi = {10.1098/rspa.1935.0122},
    url = {https://doi.org/10.1098/rspa.1935.0122},
}

@article{mezard2002analytic,
  author    = {M{\'e}zard, M. and Parisi, G. and Zecchina, R.},
  title     = {Analytic and algorithmic solution of random satisfiability problems},
  journal   = {Science},
  volume    = {297},
  number    = {5582},
  pages     = {812--815},
  year      = {2002},
  publisher = {American Association for the Advancement of Science},
  doi       = {10.1126/science.1073287}
}

@article{PhysRevE.76.045101,
  title = {Component sizes in networks with arbitrary degree distributions},
  author = {Newman, M. E. J.},
  journal = {Phys. Rev. E},
  volume = {76},
  issue = {4},
  pages = {045101(R)},
  numpages = {4},
  year = {2007},
  month = {Oct},
  publisher = {American Physical Society},
  doi = {10.1103/PhysRevE.76.045101},
  url = {https://link.aps.org/doi/10.1103/PhysRevE.76.045101}
}

@article{PhysRevE.82.016101,
  title = {Message passing approach for general epidemic models},
  author = {Karrer, Brian and Newman, M. E. J.},
  journal = {Phys. Rev. E},
  volume = {82},
  issue = {1},
  pages = {016101},
  numpages = {9},
  year = {2010},
  month = {Jul},
  publisher = {American Physical Society},
  doi = {10.1103/PhysRevE.82.016101},
  url = {https://link.aps.org/doi/10.1103/PhysRevE.82.016101}
}

@article{Newman_2023, title={Message passing methods on complex networks}, volume={479}, DOI={10.1098/rspa.2022.0774}, number={2270}, journal={Proceedings of the Royal Society A: Mathematical, Physical and Engineering Sciences}, author={Newman, M. E.}, year={2023}}

@incollection{Dorogovtsev_F._2022,
    author = {Dorogovtsev, Sergey N. and Mendes, José F. F.},
    isbn = {9780199695119},
    title = {Networks of Networks},
    booktitle = {The Nature of Complex Networks},
    publisher = {Oxford University Press},
    year = {2022},
    month = {06},
    doi = {10.1093/oso/9780199695119.003.0008},
    url = {https://doi.org/10.1093/oso/9780199695119.003.0008}
}

@book{Newman_2019, place={Oxford}, title={Networks}, publisher={Oxford University Press}, author={Newman, Mark E.J}, year={2019}}

@article{PhysRevE.107.054303,
  title = {Belief propagation on networks with cliques and chordless cycles},
  author = {Mann, Peter and Dobson, Simon},
  journal = {Phys. Rev. E},
  volume = {107},
  issue = {5},
  pages = {054303},
  numpages = {13},
  year = {2023},
  month = {May},
  publisher = {American Physical Society},
  doi = {10.1103/PhysRevE.107.054303},
  url = {https://link.aps.org/doi/10.1103/PhysRevE.107.054303}
}

@inproceedings{Yedidia2001Generalized,
  author    = {Jonathan S. Yedidia and William T. Freeman and Yair Weiss},
  title     = {Generalized Belief Propagation},
  booktitle = {Proceedings of the 14th Annual Conference on Neural Information Processing Systems},
  series    = {NIPS '00},
  editor    = {T. G. Dietterich and S. Becker and Z. Ghahramani},
  pages     = {689--695},
  publisher = {MIT Press},
  address   = {Cambridge, MA},
  year      = {2001}
}

@article{PhysRevE.111.064301,
  title = {Alternative expression of message passing on networks},
  author = {Mann, Peter and Dobson, Simon},
  journal = {Phys. Rev. E},
  volume = {111},
  issue = {6},
  pages = {064301},
  numpages = {7},
  year = {2025},
  month = {Jun},
  publisher = {American Physical Society},
  doi = {10.1103/PhysRevE.111.064301},
  url = {https://link.aps.org/doi/10.1103/PhysRevE.111.064301}
}

@article{Zhang_2012, title={Inference of Kinetic Ising model on sparse graphs}, volume={148}, DOI={10.1007/s10955-012-0547-1}, number={3}, journal={Journal of Statistical Physics}, author={Zhang, Pan}, year={2012}, month={Aug}, pages={502–512}}

@article{PhysRevX.13.031021,
  title = {Backtracking Dynamical Cavity Method},
  author = {Behrens, Freya and Hudcov\'a, Barbora and Zdeborov\'a, Lenka},
  journal = {Phys. Rev. X},
  volume = {13},
  issue = {3},
  pages = {031021},
  numpages = {14},
  year = {2023},
  month = {Aug},
  publisher = {American Physical Society},
  doi = {10.1103/PhysRevX.13.031021},
  url = {https://link.aps.org/doi/10.1103/PhysRevX.13.031021}
}

@article{Neri_2009,
doi = {10.1088/1742-5468/2009/08/P08009},
url = {https://doi.org/10.1088/1742-5468/2009/08/P08009},
year = {2009},
month = {aug},
publisher = {},
volume = {2009},
number = {08},
pages = {P08009},
author = {Neri, I and Bollé, D},
title = {The cavity approach to parallel dynamics of Ising spins on a graph},
journal = {Journal of Statistical Mechanics: Theory and Experiment}
}

@article{J_P_L_Hatchett_2004,
doi = {10.1088/0305-4470/37/24/001},
url = {https://doi.org/10.1088/0305-4470/37/24/001},
year = {2004},
month = {jun},
publisher = {},
volume = {37},
number = {24},
pages = {6201},
author = {J P L Hatchett and B Wemmenhove and I Pérez Castillo and T Nikoletopoulos and N S Skantzos and A C C Coolen},
title = {Parallel dynamics of disordered Ising spin systems on finitely connected random graphs},
journal = {Journal of Physics A: Mathematical and General}
}

@article{Mimura_2009,
doi = {10.1088/1751-8113/42/41/415001},
url = {https://doi.org/10.1088/1751-8113/42/41/415001},
year = {2009},
month = {sep},
publisher = {},
volume = {42},
number = {41},
pages = {415001},
author = {Mimura, Kazushi and Coolen, A C C},
title = {Parallel dynamics of disordered Ising spin systems on finitely connected directed random graphs with arbitrary degree distributions},
journal = {Journal of Physics A: Mathematical and Theoretical}
}

@article{kanoria2011majority,
  title={Majority dynamics on trees and the dynamic cavity method},
  author={Kanoria, Yuval and Montanari, Andrea},
  journal={The Annals of Applied Probability},
  volume={21},
  number={5},
  pages={1694--1748},
  year={2011},
  publisher={Institute of Mathematical Statistics},
  url={https://projecteuclid.org/journals/annals-of-applied-probability/volume-21/issue-5/Majority-dynamics-on-trees-and-the-dynamic-cavity-method/10.1214/10-AAP729.full}
}

@article{Mann_Smith_Mitchell_Dobson_2021, title={Symbiotic and antagonistic disease dynamics on networks using Bond percolation}, volume={104}, DOI={10.1103/physreve.104.024303}, number={2}, journal={Physical Review E}, author={Mann, Peter and Smith, V. Anne and Mitchell, John B. and Dobson, Simon}, year={2021}, month={Aug}}

@article{PhysRevE.97.010104,
  title = {Matrix product algorithm for stochastic dynamics on networks applied to nonequilibrium Glauber dynamics},
  author = {Barthel, Thomas and De Bacco, Caterina and Franz, Silvio},
  journal = {Phys. Rev. E},
  volume = {97},
  issue = {1},
  pages = {010104(R)},
  numpages = {6},
  year = {2018},
  month = {Jan},
  publisher = {American Physical Society},
  doi = {10.1103/PhysRevE.97.010104},
  url = {https://link.aps.org/doi/10.1103/PhysRevE.97.010104}
}

@article{PhysRevLett.132.117401,
  title = {Tensor Network Message Passing},
  author = {Wang, Yijia and Zhang, Yuwen Ebony and Pan, Feng and Zhang, Pan},
  journal = {Phys. Rev. Lett.},
  volume = {132},
  issue = {11},
  pages = {117401},
  numpages = {6},
  year = {2024},
  month = {Mar},
  publisher = {American Physical Society},
  doi = {10.1103/PhysRevLett.132.117401},
  url = {https://link.aps.org/doi/10.1103/PhysRevLett.132.117401}
}

@article{Mann_Smith_Mitchell_Dobson_2022, title={N-strain epidemic model using bond percolation}, volume={106}, DOI={10.1103/physreve.106.014304}, number={1}, journal={Physical Review E}, author={Mann, Peter and Smith, V. Anne and Mitchell, John B. and Dobson, Simon}, year={2022}, month={Jul}}

@article{Cantwell_Newman_2019, title={Message passing on networks with loops}, volume={116}, DOI={10.1073/pnas.1914893116}, number={47}, journal={Proceedings of the National Academy of Sciences}, author={Cantwell, George T. and Newman, M. E.}, year={2019}, pages={23398–23403}}

@article{pgxg-mhjr,
  title = {Belief propagation for finite networks using a symmetry-breaking source node},
  author = {Kim, Seongmin and Kirkley, Alec},
  journal = {Phys. Rev. Res.},
  volume = {8},
  issue = {1},
  pages = {013202},
  numpages = {6},
  year = {2026},
  month = {Feb},
  publisher = {American Physical Society},
  doi = {10.1103/pgxg-mhjr},
  url = {https://link.aps.org/doi/10.1103/pgxg-mhjr}
}

@article{PhysRevX.3.021004,
  title = {Binary-State Dynamics on Complex Networks: Pair Approximation and Beyond},
  author = {Gleeson, James P.},
  journal = {Phys. Rev. X},
  volume = {3},
  issue = {2},
  pages = {021004},
  numpages = {20},
  year = {2013},
  month = {Apr},
  publisher = {American Physical Society},
  doi = {10.1103/PhysRevX.3.021004},
  url = {https://link.aps.org/doi/10.1103/PhysRevX.3.021004}
}

@article{PhysRevLett.107.068701,
  title = {High-Accuracy Approximation of Binary-State Dynamics on Networks},
  author = {Gleeson, James P.},
  journal = {Phys. Rev. Lett.},
  volume = {107},
  issue = {6},
  pages = {068701},
  numpages = {4},
  year = {2011},
  month = {Aug},
  publisher = {American Physical Society},
  doi = {10.1103/PhysRevLett.107.068701},
  url = {https://link.aps.org/doi/10.1103/PhysRevLett.107.068701}
}

@article{Shrestha_Scarpino_Moore_2015, title={Message-passing approach for recurrent-state epidemic models on networks}, volume={92}, DOI={10.1103/physreve.92.022821}, number={2}, journal={Physical Review E}, author={Shrestha, Munik and Scarpino, Samuel V. and Moore, Cristopher}, year={2015}, month={Aug}}

@article{385j-2f29,
  title = {Threshold and quasistationary distribution for the susceptible-infectious-susceptible model on networks},
  author = {Cantwell, George T. and Moore, Cristopher},
  journal = {Phys. Rev. E},
  volume = {113},
  issue = {6},
  pages = {064305},
  numpages = {11},
  year = {2026},
  month = {Jun},
  publisher = {American Physical Society},
  doi = {10.1103/385j-2f29},
  url = {https://link.aps.org/doi/10.1103/385j-2f29}
}

@article{Harris_1974, title={Contact interactions on a lattice}, volume={2}, DOI={10.1214/aop/1176996493}, number={6}, journal={The Annals of Probability}, author={Harris, T. E.}, year={1974}, month={Dec}}

@article{Castellano_PastorSatorras_2018,
author  = {Castellano, Claudio and Pastor-Satorras, Romualdo},
title   = {Relevance of backtracking paths in recurrent-state epidemic spreading on networks},
journal = {Phys. Rev. E}, volume = {98}, pages = {052313}, year = {2018},
doi     = {10.1103/PhysRevE.98.052313}}

@article{Lokhov2015Dynamic,
  author = {Lokhov, Andrey Y. and Mézard, Marc and Zdeborová, Lenka},
  title = {Dynamic message-passing equations for models with unidirectional dynamics},
  journal = {Phys. Rev. E},
  volume = {91},
  issue = {1},
  pages = {012811},
  numpages = {15},
  year = {2015},
  month = {Jan},
  publisher = {American Physical Society},
  doi = {10.1103/PhysRevE.91.012811},
  url = {https://link.aps.org/doi/10.1103/PhysRevE.91.012811}
}

@book{gallager1963low, title={Low-Density Parity-Check Codes}, author={Gallager, Robert G.}, year={1963}, publisher={MIT Press}, address={Cambridge, MA}}

@book{jordan1998learning,
  editor    = {Michael I. Jordan},
  title     = {Learning in Graphical Models},
  publisher = {Kluwer Academic Publishers},
  address   = {Dordrecht, Netherlands},
  year      = {1998},
  series    = {NATO Science Series D: Information and Communication Sciences},
  volume    = {89},
  isbn      = {978-0792350176}
}

@inproceedings{NIPS1997_0245952e,
 author = {Frey, Brendan J and MacKay, David},
 booktitle = {Advances in Neural Information Processing Systems},
 editor = {M. Jordan and M. Kearns and S. Solla},
 pages = {},
 publisher = {MIT Press},
 title = {A Revolution: Belief Propagation in Graphs with Cycles},
 url = {https://proceedings.neurips.cc/paper_files/paper/1997/file/0245952ecff55018e2a459517fdb40e3-Paper.pdf},
 volume = {10},
 year = {1997}
}

@article{PhysRevE.108.034310,
  title = {Heterogeneous message passing for heterogeneous networks},
  author = {Cantwell, George T. and Kirkley, Alec and Radicchi, Filippo},
  journal = {Phys. Rev. E},
  volume = {108},
  issue = {3},
  pages = {034310},
  numpages = {11},
  year = {2023},
  month = {Sep},
  publisher = {American Physical Society},
  doi = {10.1103/PhysRevE.108.034310},
  url = {https://link.aps.org/doi/10.1103/PhysRevE.108.034310}
}

@article{
doi:10.1126/sciadv.abf1211,
author = {Alec Kirkley  and George T. Cantwell  and M. E. J. Newman },
title = {Belief propagation for networks with loops},
journal = {Science Advances},
volume = {7},
number = {17},
pages = {eabf1211},
year = {2021},
doi = {10.1126/sciadv.abf1211},
URL = {https://www.science.org/doi/abs/10.1126/sciadv.abf1211},
eprint = {https://www.science.org/doi/pdf/10.1126/sciadv.abf1211}}
\end{document}